\documentclass[sigconf]{acmart}
\AtBeginDocument{%
  }

\setcopyright{acmlicensed}
\copyrightyear{2018}
\acmYear{2018}
\acmDOI{XXXXXXX.XXXXXXX}
\acmConference[Conference acronym 'XX]{Make sure to enter the correct
  conference title from your rights confirmation email}{June 03--05,
  2018}{Woodstock, NY}
\acmISBN{978-1-4503-XXXX-X/2018/06}

\usepackage{enumitem}
\usepackage{pifont}
\usepackage{multirow}
\usepackage{multicol}
\usepackage{makecell}
\usepackage{adjustbox}
\usepackage{tikz}
\usepackage{pgfplots}
\usepackage{pgfplotstable}
\usepackage{float}
\usepackage{booktabs}
\usepackage{xcolor}
\usepackage{colortbl}
\usepackage{array,booktabs,multirow}
\usepackage{booktabs,multirow,array}
\usepackage{bm} 
\usepackage{graphicx}
\usepackage{subcaption}
\usepackage[table]{xcolor}
\usepackage{tabularx}
\usepackage{tabularray}
\usepackage{array}
\usepackage[most]{tcolorbox}
\definecolor{LightBlue}{RGB}{230,245,255} 
\definecolor{LightGreen}{RGB}{235,255,240}

\newcolumntype{L}[1]{>{\raggedright\arraybackslash}m{#1}}
\newcolumntype{C}[1]{>{\centering\arraybackslash}m{#1}}
\newcolumntype{Y}{>{\raggedright\arraybackslash}X}

\definecolor{UpRed}{RGB}{150,0,0}
\definecolor{DownGreen}{RGB}{0,100,0}

\newcommand{\diff}[1]{\hspace{2pt}{\scriptsize\bfseries\makebox[0.80cm][l]{(#1)}}}
\newcommand{\updiff}[1]{\diff{{\color{UpRed}$\bm{\uparrow}#1$}}}
\newcommand{\downdiff}[1]{\diff{{\color{DownGreen}$\bm{\downarrow}#1$}}}

\definecolor{boxcolor}{RGB}{46,139,87}

\tcbset{
  mygreenbox/.style={
    colback=boxcolor!10!white, 
    colframe=boxcolor!70!black, 
    boxrule=0pt,          
    arc=2pt,                
    left=6pt, right=6pt, top=6pt, bottom=6pt 
  }
}
\tcbset{
  mygraybox/.style={
    colback=black!5,      
    colframe=black!35,    
    boxrule=0pt,
    fonttitle=\Large\bfseries,
    fontupper=\large,
    arc=2pt,
    left=6pt, right=6pt, top=6pt, bottom=6pt
  }
}
\tcbset{
  mypurplebox/.style={
    colback=purple!8,     
    colframe=purple!60,   
    boxrule=0pt,
    fonttitle=\Large\bfseries,
    fontupper=\large,
    arc=2pt,
    left=6pt, right=6pt, top=6pt, bottom=6pt
  }
}
\tcbset{
  mybluebox/.style={
    colback=blue!4,                 
    colframe=blue!55!black,         
    boxrule=0pt,
    fonttitle=\Large\bfseries,
    fontupper=\large,
    arc=2pt,
    left=6pt, right=6pt, top=6pt, bottom=6pt
  }
}

\usetikzlibrary{patterns}
\begin{document}

\title{Do We Still Need GraphRAG? Benchmarking RAG and GraphRAG for Agentic Search Systems}


\author{Dongzhe Fan}
\affiliation{%
  \institution{New York University Shanghai}
  \city{Shanghai}
  \country{China}}
\email{df2362@nyu.edu}

\author{Zheyi Xue, Siyuan Liu}
\affiliation{%
  \institution{New York University Shanghai}
  \city{Shanghai}
  \country{China}
}
\email{{zx1793,sl11766}@nyu.edu}

\author{Qiaoyu Tan}
\affiliation{%
  \institution{New York University Shanghai}
  \city{Shanghai}
  \country{China}}
\email{qiaoyu.tan@nyu.edu}

\renewcommand{\shortauthors}{Trovato et al.}

\begin{abstract}
Retrieval-augmented generation (RAG) and its graph-based extensions (GraphRAG) are effective paradigms for improving large language model (LLM) reasoning by grounding generation in external knowledge. However, most existing RAG and GraphRAG systems operate under static or one-shot retrieval, where a fixed set of documents is provided to the LLM in a single pass. In contrast, recent agentic search systems enable dynamic, multi-round retrieval and sequential decision-making during inference, and have shown strong gains when combined with vanilla RAG by introducing implicit structure through interaction.
This progress raises a fundamental question: \textit{can agentic search compensate for the absence of explicit graph structure, reducing the need for costly GraphRAG pipelines?}
To answer this question, we introduce RAGSearch, a unified benchmark that evaluates dense RAG and representative GraphRAG methods as retrieval infrastructures under agentic search. RAGSearch covers both training-free and training-based agentic inference across multiple question answering benchmarks. To ensure fair and reproducible comparison, we standardize the LLM backbone, retrieval budgets, and inference protocols, and report results on full test sets. Beyond answer accuracy, we report offline preprocessing cost, online inference efficiency, and stability. Our results show that agentic search substantially improves dense RAG and narrows the performance gap to GraphRAG, particularly in RL-based settings. Nevertheless, GraphRAG remains advantageous for complex multi-hop reasoning, exhibiting more stable agentic search behavior when its offline cost is amortized. Together, these findings clarify the complementary roles of explicit graph structure and agentic search, and provide practical guidance on retrieval design for modern agentic RAG systems.
The benchmark code and evaluation scripts are publicly available at \href{https://github.com/FanDongzhe123/RAGSearch}{https://github.com/FanDongzhe123/RAGSearch}.

\end{abstract}




\maketitle

\section{Introduction}
Retrieval-augmented generation (RAG) is a widely adopted paradigm for grounding large language models (LLMs) with external knowledge by retrieving relevant documents or text chunks at inference time~\cite{lewis2021rag, jeong2024adaptiverag, wang2024maferw}. Owing to its simplicity and efficiency, dense-retrieval-based RAG has become a standard component in knowledge-intensive applications. More recently, graph-based RAG (GraphRAG) methods~\cite{edge2024local,han2024grapgrag} have been proposed to further improve reasoning performance by explicitly organizing retrieved content into structured representations—such as hierarchical trees~\cite{edge2024local,sarthi2024raptor},, entity graphs~\cite{he2024gretriever, gutiérrez2025hipporag, guo2025lightrag, gutiérrez2025hipporag2}, or hypergraphs~\cite{luo2025hypergraphrag, feng2025hyperrag}, enabling more effective multi-hop reasoning and evidence aggregation.
\begin{figure}[htp]
\centering
\includegraphics[width=8.5cm, height=6cm]{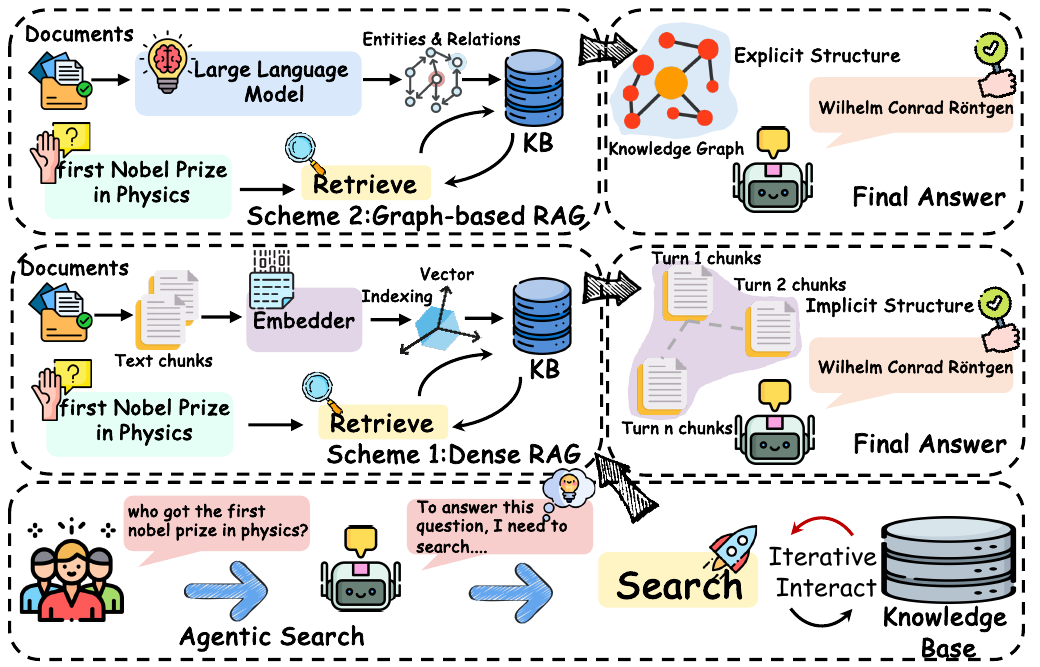}
\caption{Explicit vs. Implicit Structure in RAG Systems. GraphRAG relies on explicit graph construction, whereas agentic search over dense RAG can induce implicit evidence structure through multi-round retrieval and reasoning.}
\label{fig:demo}
\end{figure}

Despite their effectiveness, existing RAG and GraphRAG systems are predominantly designed for a static or one-shot retrieval setting~\cite{lee2025hybgrag, lewis2021rag, chen2025pathragpruninggraphbasedretrieval}, where a fixed set of retrieved documents is provided to the LLM in a single pass. This assumption limits the ability of retrieval to adapt to intermediate reasoning states during inference. In parallel, agentic search systems~\cite{yao2022react,openai2024openaio1, li2025searcho1, jin2025searchr1, liu2026graphsearch, luo2025kbqao1agenticknowledgebase} have recently emerged as a powerful alternative, shifting retrieval from a static preprocessing step to a dynamic, multi-round process. By enabling sequential decision-making, agentic systems allow LLMs to iteratively refine queries and retrieval strategies based on partial reasoning outcomes, and have demonstrated strong empirical gains when combined with vanilla RAG~\cite{lewis2021rag}.

These developments expose a fundamental tension in modern retrieval-augmented systems. From a structural perspective (Figure \ref{fig:demo}), dense RAG retrieves and ranks text chunks independently based on semantic similarity, without explicitly modeling relationships among retrieved units. GraphRAG introduces explicit graph structure, injecting relational inductive bias to guide multi-step retrieval and reasoning. Agentic search, in contrast, introduces implicit structure through interaction, using sequential control to organize information access during inference. This raises a central question:

\textit{Can agentic search compensate for the lack of explicit graph structure in dense RAG, or does GraphRAG remain necessary under agentic inference?}

Answering this question is non-trivial. While prior work shows that GraphRAG can outperform dense RAG on multi-hop and compositional reasoning tasks, these gains come with substantial offline preprocessing cost, including entity extraction, summarization, graph construction, and index maintenance. Whether such costs are justified becomes increasingly unclear in the presence of agentic search, which may reduce reliance on explicit structure by enabling deeper exploration and iterative refinement at inference time.

However, despite the importance of this question, existing evaluations do not provide a definitive answer. Prior studies~\cite{luo2025graphr1,jin2025searchr1,lee2025hybgrag,yu2025graphragr1} often adopt inconsistent evaluation protocols, rely on partial test sets, or vary computational budgets across methods—particularly for agentic systems, where retrieval calls and token usage are rarely controlled. Moreover, GraphRAG methods are typically evaluated as monolithic end-to-end systems~\cite{dong2025youtu}, rather than as retrieval infrastructures that can be reused across different inference paradigms. These limitations make it difficult to assess when explicit graph structure is truly necessary and when agentic search can serve as an effective substitute.

To fill in this gap, we introduce RAGSearch, a unified benchmark for studying retrieval-augmented generation under agentic search. RAGSearch treats dense RAG and graph-based RAG as alternative retrieval infrastructures within the RAG paradigm, and evaluates their interaction with agentic inference under unified protocols, matched retrieval budgets, and full test-set evaluation. Our \textbf{key contribution}s are as follows:
\vspace{-10pt}

\begin{itemize}[noitemsep,leftmargin=*]
\item \textbf{A unified benchmark.} We introduce RAGSearch, the first and timely benchmark that systematically evaluates dense RAG and multiple representative GraphRAG pipelines as retrieval infrastructures under agentic search systems. RAGSearch unifies datasets, LLM backbones, retrieval budgets, and evaluation protocols, enabling controlled comparison across static retrieval and dynamic, agentic inference settings.
\item \textbf{A comprehensive evaluation.} We implement and benchmark both training-free agentic search methods and reinforcement-learning–optimized agentic systems on top of vanilla RAG and five representative GraphRAG methods. This allows us to study how different forms of retrieval structure interact with agentic control across inference paradigms, rather than evaluating retrieval or agents in isolation. 
\item \textbf{Multi-dimensional analysis.} Beyond answer accuracy, RAGSearch reports offline preprocessing cost, online inference efficiency, and stability under agentic control, revealing when agentic search can compensate for structure-agnostic retrieval and when graph-based retrieval remains beneficial despite higher construction cost. All code, configurations, and leaderboards are released to support reproducible evaluation and future extensions.
\end{itemize}

\section{Related Work}
Our work is closely related to the following three directions.

\noindent \textbf{RAG-based LLM reasoning.}
Large language models (LLMs) have made remarkable progress, yet they still have notable limitations, particularly in domain-specific or knowledge-intensive scenarios, which often generate hallucinated content when queries go beyond their training data or require up-to-date information. To mitigate these issues, Retrieval-Augmented Generation (RAG)~\cite{lewis2021rag} augments LLMs by retrieving relevant document chunks from an external knowledge base. Building on this pipeline, recent work on RAG-based LLM reasoning focuses on improving how models plan, retrieve, and reason over external evidence. Broadly, existing approaches can be categorized into train-free~\cite{he2024retrievingrethinkingrevisingchainofverification, zhang2025arise} and train-based paradigms~\cite{wang2025kblam, islam2024openrag, zhang2024raftadaptinglanguagemodel}. 

\noindent \textbf{GraphRAG-enhanced LLM reasoning.}
Although standard RAG effectively grounds LLMs with retrieved textual evidence, it may fall short for multi-hop or relational queries where supporting information is distributed across multiple documents. GraphRAG~\cite{edge2024local} addresses these limitations by leveraging graph-structured representations of knowledge. Motivated by this formulation, recent works have expanded the design space of GraphRAG and proposed more computationally efficient graph construction pipelines. RAPTOR~\cite{sarthi2024raptor} constructs a hierarchical tree index by recursively clustering and summarizing text chunks, enabling retrieval at multiple levels of abstraction for improved long-context and multi-hop reasoning. Inspired by the hippocampal memory indexing theory, HippoRAG2~\cite{gutiérrez2025hipporag2} constructs an entity-centric corpus graph from extracted facts and applies Personalized PageRank–style propagation to retrieve multi-hop, relation-aware evidence across documents. Going beyond traditional graph-structured representations, HyperGraphRAG~\cite{luo2025hypergraphrag} constructs the knowledge base as a hypergraph to capture higher-order relations. While LinearRAG~\cite{zhuang2025linearrag} constructs a relation-free hierarchical "Tri-Graph" using lightweight entity extraction and semantic linking, enabling scalable graph-based retrieval without costly relation extraction.

\noindent \textbf{Agentic Search.}
However, most existing GraphRAG systems still adhere to a single-shot retrieval paradigm, recent methods~\cite{li2025searcho1, liu2026graphsearch, luo2025graphr1, jin2025searchr1} explore multi-step retrieval by iteratively refining queries under LLM guidance. Search-o1~\cite{li2025searcho1} interleaves step-by-step LLM reasoning with on-demand external retrieval, and uses a Reason-in-Documents module to refine retrieved documents before integrating them into the reasoning process. GraphSearch~\cite{liu2026graphsearch} enables iterative multi-step retrieval by jointly querying textual chunks and GraphRAG for complex multi-hop reasoning. Besides the training-free approaches, recent work also explores train-based methods. Search-R1~\cite{jin2025searchr1} adopts an RL-based training paradigm to teach an LLM to interleave step-by-step reasoning with multi-turn search, typically instantiated with dense retrieval over an external corpus to provide information for each reasoning step. While Graph-R1~\cite{luo2025graphr1} extends multi-turn search to the GraphRAG settings.
\begin{figure*}[ht!]
\centering
\includegraphics[width=15.5cm, height=6.5cm]{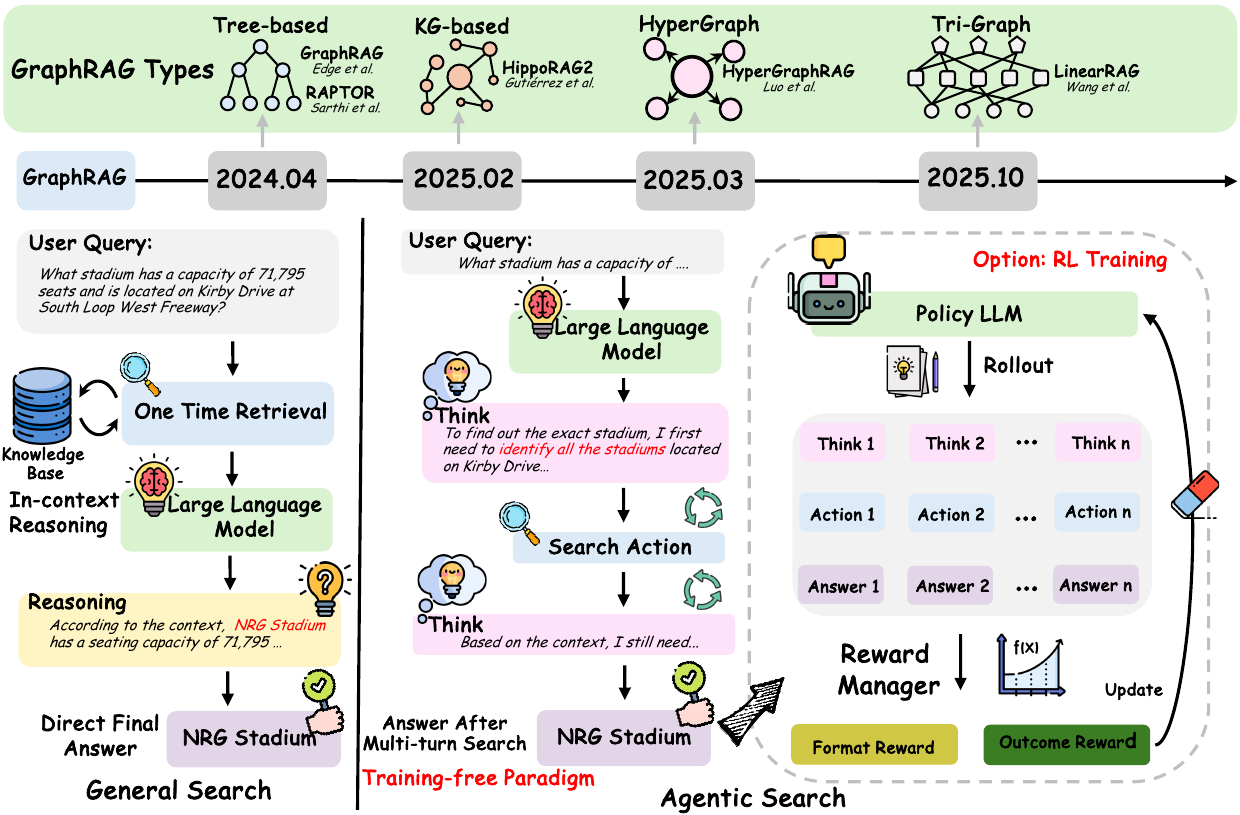}
\caption{\textbf{Overview of the RAGSearch benchmark. RAGSearch models agentic search as an LLM agent interacting with interchangeable retrieval backends (dense RAG or GraphRAG) through a unified interface, and benchmarks both training-free agentic search and RL-based agentic search under standardized protocols. 
}}
\label{fig:main}
\end{figure*}

\section{Preliminary}




We consider open-domain question answering with an input query $q$ and a large language model (LLM) $\mathcal{M}$. Under standard LLM reasoning, the model generates an answer $y$ conditioned solely on the query: $y \sim \mathcal{M}(q).$
While powerful, this formulation relies entirely on the knowledge encoded in model parameters and is limited in its ability to handle knowledge-intensive or multi-hop reasoning tasks that require access to external information.

\noindent\textbf{Retrieval-Augmented Reasoning (RAG).}
RAG extends LLM reasoning by grounding generation in an external knowledge corpus $\mathcal{K}=\{d_1,\dots,d_N\}$. A RAG system consists of a retriever $\mathcal{R}$ and an LLM $\mathcal{M}$. Given a query $q$, the retriever selects a set of relevant documents or text chunks $C_q = \mathcal{R}(q \mid \mathcal{K})$,
which are appended to the model input for answer generation: $y \sim \mathcal{M}(q, C_q).$
In most existing RAG pipelines, retrieval is performed once prior to decoding, and the retrieved context remains fixed throughout generation. This \emph{static} or \emph{one-shot retrieval} assumption underlies the majority of dense-retrieval-based RAG systems.

\noindent\textbf{Graph-Based Retrieval-Augmented Generation (GraphRAG)}
GraphRAG further extends retrieval-augmented reasoning by explicitly organizing the knowledge corpus into a structured graph or hypergraph prior to inference. We abstract a GraphRAG knowledge base as $\mathcal{G} = (\mathcal{V}, \mathcal{E})$,
where nodes $\mathcal{V}$ represent textual units (e.g., documents, entities, or summaries) and edges $\mathcal{E}$ encode relationships among them. Retrieval in GraphRAG typically selects nodes, paths, or subgraphs from $\mathcal{G}$, which are then provided as structured evidence to the LLM: $y \sim \mathcal{M}(q, Z_q), \quad Z_q \subseteq \mathcal{G}$. Compared to dense RAG, GraphRAG introduces explicit structural inductive bias that can improve multi-hop reasoning and evidence aggregation. However, similar to standard RAG, GraphRAG retrieval is typically in a single preprocessing step, and the selected evidence is appended to the model input before decoding.

In this work, we focus on benchmarking RAG and GraphRAG under \emph{agentic search} settings, where retrieval is integrated into inference and performed iteratively during decoding. This paradigm enables multi-round retrieval and adaptive information access based on intermediate reasoning states, fundamentally changing how retrieval interacts with LLM reasoning.

\section{RAGSearch Benchmark}

In this section, we introduce \textbf{RAGSearch}, a benchmark designed to systematically study how different retrieval infrastructures interact with agentic search systems. RAGSearch treats dense RAG and GraphRAG as interchangeable \emph{retrieval backends} within a unified agentic search framework, enabling controlled and fair comparison across agentic inference paradigms under standardized settings. An overview of the RAGSearch framework is shown in Figure~\ref{fig:main}.
We first formalize a general agentic search abstraction and its interaction with retrieval backends in Section~4.1. We then describe two representative agentic pipelines instantiated in RAGSearch: \emph{training-free agentic search}, which relies on structured prompting and heuristic control (Section~4.2), and \emph{reinforcement-learning–based agentic search}, where the agent policy is optimized using domain-specific data (Section~4.3). Finally, Section~4.4 specifies the retrieval backends included in RAGSearch and how they are instantiated as environments within the benchmark.

\subsection{General Agentic Search Formulation}
\label{sec:agentic_formulation}

We formalize agentic search in RAGSearch using a high-level abstraction inspired by the ReAct~\cite{yao2022react} framework, which models inference as an interleaved loop of \emph{reasoning} and \emph{interaction} with a retrieval environment. Given an input query $q$, an agent equipped with an LLM $\mathcal{M}$ interacts with a retrieval backend $\mathcal{B}$ (e.g., dense RAG or GraphRAG) over multiple steps.

At each step $t$, the agent performs reasoning delimited by
\texttt{\textcolor{blue}{<think>}} and \texttt{\textcolor{blue}{</think>}},
conditioning on the available information and the interaction history, to decide
whether to invoke retrieval or to terminate and produce a final answer.
When retrieval is triggered, the agent emits a search query $q_t$ demarcated by
\texttt{\textcolor{orange}{<search>}} and \texttt{\textcolor{orange}{</search>}}.
The system then executes a retrieval operation over $\mathcal{B}$:
\begin{equation}
    \mathcal{I}^{q}_{t} = \textsc{Retrieve}(q_t, \mathcal{B}).
\end{equation}
Here, $\mathcal{I}^{q}_{t}$ denotes the retrieved information, which can be a set of
text chunks $C_{q_t}$ or a subgraph $Z_{q_t}$. The information is wrapped with
\texttt{\textcolor{red}{<information>}} and \texttt{\textcolor{red}{</information>}}
and appended to the ongoing reasoning sequence. This process continues iteratively until the agent decides to produce the final answer within \texttt{\textcolor{green}{<answer>}} and \texttt{\textcolor{green}{</answer>}}. The generated reasoning process can be expressed as:
\begin{equation}
\begin{aligned}
P(R,a \mid \mathcal{P},\mathcal{B})
&=
\underbrace{\prod_{t=1}^{T_{R}} P\!\left(R_t \mid R_{<t}, \mathcal{P},q_t, \mathcal{I}^{q}_{<t}\right)}_{\text{\textcolor{blue}{Reasoning Process}}}
\cdot
\underbrace{\prod_{t=1}^{T_a} P\!\left(a \mid R, \mathcal{I}^q,q\right)}_{\text{\textcolor{green}{Answer Generation}}} \, 
\end{aligned}
\end{equation}
Where $\mathcal{P}$ indicates the system template, $R_t$ is the generated reasoning sequence at step $t$.

This formulation intentionally abstracts away low-level details of action spaces and memory representations and highlights two key properties shared by modern agentic systems: (i) retrieval is performed \emph{dynamically during inference}, rather than as a one-shot preprocessing step; and (ii) the same agentic control logic can operate over different retrieval infrastructures. RAGSearch adopts this abstraction to decouple agentic reasoning from retrieval backends, enabling systematic benchmarking of dense RAG and GraphRAG under a unified agentic search framework.



\subsection{Training-Free Agentic Search Pipelines}
\label{sec:training_free_agents}
Based on the general agentic search formulation in Section~\ref{sec:agentic_formulation}, we instantiate a class of \emph{training-free agentic search pipelines} that reflect the design of state-of-the-art systems such as Search-o1~\cite{li2025searcho1} and GraphSearch~\cite{yang2025graphsearch}. These methods do not learn an explicit control policy; instead, they rely on structured prompting and heuristic control to guide multi-round retrieval and reasoning during inference. Concretely, contemporary training-free agentic search systems often fall into two representative paradigms: (i) \textbf{reasoning-driven on-demand search}, where the model triggers retrieval adaptively based on uncertainty or information needs (e.g., Search-o1), and (ii) \textbf{orchestrated multi-agent workflows}, where modular roles are explicitly coordinated via prompting and routing to perform decomposition, verification, and iterative retrieval (e.g., GraphSearch).
\subsubsection{Reasoning-driven On-demand Search}
This paradigm typically relies on the model’s own reasoning process to decide whether to invoke search. Compared to the basic ReAct-style loop, this paradigm augments the reasoning--interaction with Reason-in-Documents components:
\[
\texttt{Query} \rightarrow
\underbrace{\texttt{Think} \rightarrow \texttt{Search} \rightarrow \texttt{Knowledge Refinement}}_{\texttt{Iterative}}
\rightarrow \texttt{Answer}
\]
\paragraph{\textbf{Knowledge Refinement}.}
To support long-horizon reasoning without exceeding context limits, training-free agents typically summarize retrieved observations before incorporating them into the agent state. Following the design of Search-o1~\cite{li2025searcho1}, each retrieval result is compressed into a concise summary that captures the most salient evidence relevant to the current reasoning goal. 
\subsubsection{Orchestrated Multi-agent Workflows}
This paradigm foregrounds coordinated multi-module interaction, decomposing the original query into smaller, more tractable sub-queries that can be solved sequentially to answer the original question. This paradigm typically exhibits the following proceudre:
\[
\texttt{Query} \rightarrow
\underbrace{\texttt{Docomposition} \rightarrow \texttt{Search} \rightarrow \texttt{Verification}}_{\texttt{Iterative}}
\rightarrow \texttt{Answer}
\]
\paragraph{\textbf{Query Decomposition}.}
Instead of issuing a single retrieval query, the agent may decompose the original question into a sequence of sub-queries,  each representing a smaller and tractable component of the original question:
\begin{equation}
    \{q_1, q_2,...,q_i\} = Decoposition(q)
\end{equation}
Where each sub-query $q_i$ focuses on resolving a single entity, relation, or contextual dependency. Each sub-query $q_i$ and its associated retrieved documents $I_{q_i}$ are processed by the Logic Drafting module to construct a reasoning chain $\mathcal{L}$ that resolves the original problem.This mechanism, inspired by GraphSearch~\cite{yang2025graphsearch}, enables retriever to access fine-grained evidence and reducing the reasoning complexity.
\paragraph{\textbf{Evidence Verification}.}
 This module evaluates whether the accumulated evidence in $\mathcal{L}$ is sufficient and logically consistent to support a final answer, by considering factual grounding, coherence, and potential contradictions. When evidence is missing or inconsistent, the module further expands the set of sub-queries and iteratively performs additional retrieval to gather relevant context.

\noindent\textbf{Remark.}
Under this formulation, Search-o1 and GraphSearch can be viewed as different instantiations of the same training-free agentic framework, differing primarily in the retrieval environment they interact with. Dense RAG environments return unstructured text chunks, while GraphRAG environments return structured graph-based evidence. Crucially, the agentic control remains unchanged. RAGSearch adopts this unified view to benchmark training-free agentic search pipelines over both dense RAG and multiple GraphRAG infrastructures under identical inference protocols.

\subsection{RL-Based Agentic Search Training}
\label{sec:rl_agents}

While training-free agentic pipelines provide a strong and flexible baseline, their control logic is entirely prompt-driven and fixed at inference time. As a result, such agents rely heavily on the intrinsic reasoning capability of the underlying LLM backbone and cannot adapt their retrieval strategies to specific task distributions or reasoning patterns. This limitation motivates \emph{learning} the agent’s control policy from domain-specific data, enabling retrieval and reasoning behaviors to be adapted to the target task.

\paragraph{\textbf{RL-Based Agentic Search Formulation}.}
Following Search-R1 and Graph-R1, we formulate agentic search as a reinforcement learning problem, where the agent policy $\pi_\theta$ is parameterized by an LLM with trainable parameters $\theta$. Given an input query $q$, the agent interacts with a retrieval environment $\mathcal{B}$ (dense RAG or GraphRAG) and generates a trajectory
\[
\tau^{(i)} = (s^{(i)}_1, a^{(i)}_1, \dots, s^{(i)}_T, a^{(i)}_T),
\]
where each action $a_t$ corresponds to a retrieval-related decision or a termination action producing a final answer $y$. A scalar reward $r(q, y, \tau)$ is assigned to the entire trajectory after termination. The learning objective is to maximize the expected reward over the training distribution:
\begin{equation}
\max_{\pi_\theta} \;
\mathbb{E}_{q \sim \mathcal{D},\, \tau \sim \pi_\theta(\cdot \mid q, \mathcal{B})}
\big[ r(q, y, \tau) \big].
\end{equation}
This formulation directly optimizes sequence-level agent behavior, allowing the policy to learn task-specific retrieval and reasoning patterns.

To optimize the agent policy, RAGSearch adopts \emph{Group Relative Policy Optimization} (GRPO), as used in Search-R1 and Graph-R1. For each training query $q$, the agent samples a group of $K$ trajectories $\{\tau^{(1)}, \dots, \tau^{(K)}\}$ under the current policy. Each trajectory receives a reward $r^{(k)}$, which is normalized within the group to compute a relative advantage:
\[
\hat{A}(\tau_i)
=
\frac{
r^{(i)} - \operatorname{mean}\!\left(\left\{ r^{(j)}\right\}_{j=1}^{K}\right)
}{
F_{\mathrm{norm}}\!\left(\left\{ r^{(j)}\right\}_{j=1}^{K}\right)
}.
\]
The policy is then updated by increasing the likelihood of trajectories with positive relative advantage, while constraining the update via KL regularization toward a fixed reference policy $\pi_{\mathrm{ref}}$:
\begin{equation}
\begin{aligned}
\mathcal{L}_{\mathrm{GRPO}}(\theta) = \left[
\frac{1}{K}\sum_{i=1}^{K}\frac{1}{|\tau_i|}
\sum_{t=1}^{|\tau_i|}
\min\!\Big(
\rho_{\theta}\big(a^{(i)}_{t}\big)\,\hat{A}(\tau_i),g\big(\epsilon, \hat{A}(\tau_i)\big)
\Big)
\right. \\
\left.
-\;\beta\,\mathbb{D}_{\mathrm{KL}}\!\big(\pi_{\theta}\,\|\,\pi_{\mathrm{ref}}\big)
\right]
\end{aligned}
\end{equation}
where $\rho_{\theta}(a^{(i)}_{t}) = \frac{\pi_{\theta}(a_t^{(i)}|s_{t-1}^{(i)};\mathcal{B})}{\pi_{\theta_{old}}(a_t^{(i)}|s_{t-1}^{(i)};\mathcal{B})}$, $g\big(\epsilon, \hat{A}(t_i)\big)=\operatorname{clip(\rho_{\theta}(a^{(i)}_{t}), 1 \pm \epsilon)\hat{A}(\tau_i)}$, and
$\beta$ controls the strength of regularization. GRPO avoids explicit value-function learning and has been shown to be effective for training long-horizon LLM-based agents.

\paragraph{\textbf{Reward Design}.}
In RAGSearch, rewards are defined at the trajectory level and focus on task correctness and output validity. Specifically, we combine (i) an \emph{outcome-based reward} that measures answer correctness (e.g., exact match or task-specific accuracy), and (ii) a \emph{format-based reward} that encourages the agent to follow the expected interaction and answer format. Importantly, the same reward formulation is applied across dense RAG and GraphRAG retrieval environments, ensuring that learned differences in agent behavior reflect the interaction between the policy and the retrieval infrastructure rather than differences in reward design.

\noindent\textbf{Remark.}
From a unified perspective, RL-based agentic search can be viewed as learning an adaptive control policy within the same agentic framework introduced in Sections~\ref{sec:agentic_formulation} and~\ref{sec:training_free_agents}. The retrieval backends and interaction protocol remain unchanged; only the agent’s policy is optimized using domain-specific supervision. This allows RAGSearch to systematically compare training-free and learned agentic control over identical RAG and GraphRAG infrastructures, and to assess how policy learning interacts with explicit graph structure under agentic search.

\subsection{Retrieval Backends in RAGSearch}
\label{sec:retrieval_backends}

RAGSearch instantiates a fixed set of retrieval backends as interchangeable environments for agentic search. All backends expose a unified retrieval interface to the agent and differ only in how external knowledge is organized and accessed, enabling controlled comparison across retrieval infrastructures.

We include one structure-agnostic dense RAG baseline, which indexes the corpus as unstructured text chunks and retrieves evidence via semantic similarity search. In addition, we consider five representative GraphRAG backends that span diverse graph construction and retrieval strategies:
\begin{itemize}[leftmargin=*]
\item \textbf{\textit{Tree Based:}} \textbf{GraphRAG}~\cite{edge2024local}, based on hierarchical communities for multi-hop evidence aggregation; \textbf{RAPTOR}~\cite{sarthi2024raptor}, which retrieves evidence from a recursive summarization tree;
\item \textbf{\textit{Entity Graph Based:}} \textbf{HippoRAG2}~\cite{gutiérrez2025hipporag2}, which employs entity-centric graph representations;
\item \textbf{\textit{HyperGraph Based:}}   \textbf{HypergraphRAG}~\cite{luo2025hypergraphrag}, which captures higher-order relations via hyperedges;
\item \textbf{\textit{Tri-Graph Based:}} \textbf{LinearRAG}~\cite{zhuang2025linearrag}, which imposes a lightweight linear structure over retrieved content.
\end{itemize}

All GraphRAG backends incur offline preprocessing to construct their graph representations but are accessed through the same agentic interaction protocol described in Section~4.1, without backend-specific agent modifications. This standardized setup enables direct comparison between dense and graph-structured retrieval under both training-free and RL-based agentic inference.
\begin{table*}[t]
\centering
\small
\setlength{\tabcolsep}{2pt}
\renewcommand{\arraystretch}{1.10}
\caption{Overall Contain Exact Match (EM) for single-shot inference, training-free agentic systems, and RL-based agentic systems. $\spadesuit$ denotes the best result among the five GraphRAG variants. \textcolor{UpRed}{$\uparrow$} and \textcolor{DownGreen}{$\downarrow$} indicate, for the same method, the performance difference between graph-based and dense-retrieval-based RAG;$^\dagger$ denotes the in-domain dataset, and $^\star$ denotes the cross-domain dataset.}
\label{tab:1}

\begin{tabular}{@{} L{2.2cm} L{3.8cm} *{6}{C{1.5cm}} @{}}
\toprule
\multirow{2}{*}{\textbf{System}} &
\multirow{2}{*}{\textbf{Methods}} &
\multicolumn{3}{c}{\textbf{General QA}} &
\multicolumn{3}{c}{\textbf{Multi-Hop QA}} \\
\cmidrule(r){3-5}\cmidrule(l){6-8}
& & \textbf{NQ$^\dagger$} & \textbf{PopQA$^\star$} & \textbf{TriviaQA$^\star$}
  & \textbf{HotpotQA$^\dagger$} & \textbf{2Wiki$^\star$} & \textbf{Musique$^\star$} \\
\midrule

\multirow{2}{*}{\textbf{Single-shot}} &
Qwen-2.5-7B-Dense & 46.62 & 32.14 & 58.60 & 19.00 & 35.53 & 20.99 \\
& Qwen-2.5-7B-GraphRAG$^\spadesuit$ &
48.31\updiff{1.69} & 32.82\updiff{0.68} & 57.65\downdiff{0.95} &
46.70\updiff{27.70} & 62.56\updiff{27.03} & 47.95\updiff{26.96} \\
\midrule

\multirow{4}{*}{\textbf{Training-free}} &
Search-o1-7B-Dense & 38.20 & 25.57 & 58.74 & 33.76 & 29.64 & 12.62 \\
& Search-o1-7B-GraphRAG$^\spadesuit$ &
38.34\updiff{0.14} & 28.01\updiff{2.44} & 59.50\updiff{0.76} &
42.75\updiff{8.99} & 65.56\updiff{35.92} & 32.44\updiff{19.82} \\
& GraphSearch-7B-Dense & 58.27 & 36.29 & 68.70 & 38.22 & 47.43 & 13.33 \\
& GraphSearch-7B-GraphRAG$^\spadesuit$ &
61.22\updiff{2.95} & 44.77\updiff{8.48} & 72.47\updiff{3.77} &
58.64\updiff{20.42} & 79.88\updiff{32.45} & 55.26\updiff{41.93} \\
\midrule

\multirow{2}{*}{\textbf{RL-based}} &
Search-R1-7B & 48.72 & 33.10 & 63.96 & 35.76 & 33.56 & 14.42 \\
& Graph-R1-7B$^\spadesuit$ &
46.71\downdiff{2.01} & 36.23\updiff{3.13} & 66.21\updiff{2.25} &
53.42\updiff{17.66} & 66.25\updiff{32.69} & 40.82\updiff{26.40} \\
\bottomrule
\end{tabular}
\end{table*}

\section{Experiments}
In this section, we conduct extensive experiments under our benchmark setting, aiming to address the following research questions: \textbf{RQ1:} Can agentic search compensate for the absence of explicit graph structure in dense RAG? \textbf{RQ2:} Does explicit graph structure continue to provide benefits under training-free agentic search? \textbf{RQ3:} How does policy learning via reinforcement learning interact with different retrieval infrastructures? \textbf{RQ4:} How do RAG and GraphRAG differ in their robustness and stability under agentic inference? \textbf{RQ5:} What impact do different modules have in agentic systems?


\subsection{Experimental Setup}
\subsubsection{\textbf{Datasets}}
To comprehensively evaluate six standard Question Answering (QA) datasets~\cite{FlashRAG}: (1) \textbf{General QA}: Natural Questions (NQ)~\cite{nq}, PopQA~\cite{popqa} and TriviaQA~\cite{triviaqa}; (2) \textbf{Multi-hop QA}: HotpotQA~\cite{hotpotqa}, Musique~\cite{musique} and 2WikiMultiHopQA (2Wiki)~\cite{2wiki}. More details are in Appendix~\ref{app:a}.

\subsubsection{\textbf{Search Agent and RAG Systems}}
To investigate whether GraphRAG remains necessary under agentic inference, we adopt four representative agentic search systems: two training-free approaches, Search-o1~\cite{li2025searcho1} and GraphSearch~\cite{yang2025graphsearch}, and two RL-based approaches, Search-R1~\cite{jin2025searchr1} and Graph-R1~\cite{luo2025graphr1}. We also employ static one-shot retrieval methods as baselines. For dense RAG, we utilize vanilla RAG as the retrieval backend and the 2018 Wikipedia dump~\cite{wikidump} as the knowledge source. For GraphRAG, we select five representative methods: GraphRAG~\cite{edge2024local}, RAPTOR~\cite{sarthi2024raptor}, HippoRAG2~\cite{gutiérrez2025hipporag2}, HyperGraphRAG~\cite{luo2025hypergraphrag}, LinearRAG~\cite{zhuang2025linearrag}. For both training-free and RL-based agentic systems, we implemented variants using each of the six retrieval backends described above.

\subsubsection{\textbf{Implementation Details}}
We use GPT-4o-mini for knowledge construction in different graph-based retrieval infrastructures. All the training-free agent systems are implemented with Qwen2.5-7B-Instruct and Qwen2.5-32B-Instruct~\cite{qwen2025qwen25technicalreport} as LLM backbone. For the RL-based agentic systems, we adpot GRPO~\cite{shao2024grpo} as the training procedure. We utilize Qwen2.5-3B-Instruct and Qwen2.5-7B-Instruct as their LLM backbone. We jointly pre-train all the RL-based systems on HotpotQA and NQ. We randomly sample 5000 nodes from the train set of these datasets. For the test set, we utilize the full set from test set or dev set.  All experiments are done on 2 NVIDIA A100 GPUs (80GB). More details are in Appendix~\ref{app:b}.

\subsubsection{\textbf{Evaluation Metrics}}
We evaluate all the agentic systems with two metrics: F-1 and Contain Exact Match (Contain EM). More details are in Appendix~\ref{app:eval_metric}

\subsection{Overall Comparison (RQ1)}
In this section, we comprehensively compare graph-based and dense-retrieval-based RAG across different agentic systems. 
\subsubsection{Single-shot Inference}
\textbf{\textit{\underline{Observation 1}}} \textit{Under single-shot inference, dense RAG is already effective for general QA, while GraphRAG provides decisive gains primarily on multi-hop QA. } As shown in Table~\ref{tab:1}, vanilla dense RAG achieves competitive performance on general QA benchmarks, and GraphRAG yields only marginal improvements in this setting, with an average gain of +0.47. In contrast, GraphRAG substantially outperforms dense retrieval on multi-hop QA, delivering an average improvement of +27.23 across HotpotQA, 2Wiki, and Musique. This stark contrast indicates that explicit graph-structured representations are particularly beneficial for tasks requiring multi-hop evidence aggregation and compositional reasoning, while offering limited advantage for single-hop or factoid-style questions.
\subsubsection{Training-free Search Agent}
\textbf{\textit{\underline{Observation 2}}}
\textit{Agentic search can strengthen dense RAG and partially narrow the gap to GraphRAG, though the effect depends on the agent design.} Comparing Tables~\ref{tab:1} and~\ref{tab:2}, we find that the benefit of agentic search for dense RAG is not uniform. Under Search-o1, dense RAG shows mixed behavior and even performance drops on several benchmarks, indicating that generic multi-turn interaction alone does not reliably improve dense retrieval. In contrast, under GraphSearch, dense RAG improves substantially over single-shot inference across general and multi-hop QAs excluding Musique due to effective query decomposition and iterative retrieval. Quantitatively, the averaged Dense–GraphRAG gap on multi-hop QAs decreases from +27.23 (single-shot) to +26.59, and is reduced by 32.3\% relative to the second-best GraphRAG variant, suggesting that structured agentic search can partially compensate for the lack of explicit graph structure.

\begin{table*}[t]
\centering
\small
\setlength{\tabcolsep}{2pt}
\renewcommand{\arraystretch}{1.15}
\caption{Overall Contain EM results of training-free agentic systems across different retriever backends. \raisebox{0.2ex}{\colorbox{LightBlue}{\phantom{\rule{1.5em}{0.6ex}}}} and \raisebox{0.2ex}{\colorbox{LightGreen}{\phantom{\rule{1.5em}{0.6ex}}}} indicate the best and second-best results within each method, respectively. \includegraphics[height=1em]{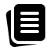} means dense-RAG, \includegraphics[height=1em]{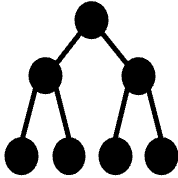} means tree-based GraphRAG, \includegraphics[height=1em]{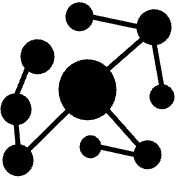} means entity graph based GraphRAG,\includegraphics[height=1em]{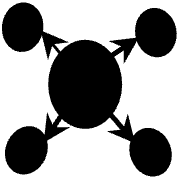} means hypergraph based GraphRAG and \includegraphics[height=1em]{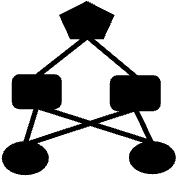} means tri-graph based GraphRAG. Avg. Rank is computed within each method (lower is better).}
\label{tab:2}

\begin{tabular}{@{} L{1.8cm} L{2.2cm} *{3}{C{1.5cm}} C{1.4cm} *{3}{C{1.5cm}} C{1.4cm} @{}}
\specialrule{1.2pt}{0pt}{0pt}  

\multirow{2}{*}{\textbf{Method}} &
\multirow{2}{*}{\textbf{Knowledge Base}} &
\multicolumn{3}{c}{\textbf{General QA}} &
\multirow{2}{*}{\textbf{Avg Rank}} &
\multicolumn{3}{c}{\textbf{Multi-Hop QA}} &
\multirow{2}{*}{\textbf{Avg Rank}} \\
\cmidrule(r){3-5}\cmidrule(l){7-9}
& & \textbf{NQ} & \textbf{PopQA} & \textbf{TriviaQA}
  & & \textbf{HotpotQA} & \textbf{2Wiki} & \textbf{Musique} & \\

\specialrule{0.9pt}{0pt}{0pt}  

\multirow{6}{*}{\textbf{Search-o1}} &
\includegraphics[height=1em]{Figures/doc.png}Dense &
\cellcolor{LightGreen}\underline{38.20} & 25.78 & \cellcolor{LightGreen}\underline{58.74} & 2.33 &
33.76 & 29.64 & 12.62 & 5.33 \\
& \includegraphics[height=1em]{Figures/hyper.png}HypergraphRAG &
33.02 & 25.57 & 56.72 & 5.33 &
33.90 & 50.58 & 28.05 & 3.67 \\
& \includegraphics[height=1em]{Figures/graph.png}HippoRAG2 &
\cellcolor{LightBlue}\textbf{38.34} & \cellcolor{LightBlue}\textbf{28.01} & \cellcolor{LightBlue}\textbf{59.50} & 1.00 &
\cellcolor{LightBlue}\textbf{42.75} & \cellcolor{LightBlue}\textbf{65.56} & \cellcolor{LightBlue}\textbf{32.44} & 1.00 \\
& \includegraphics[height=1em]{Figures/linear.png}LinearRAG &
34.32 & 25.69 & 57.03 & 4.00 &
\cellcolor{LightGreen}\underline{35.76} & \cellcolor{LightGreen}\underline{58.94} & 29.46 & 2.33 \\
& \includegraphics[height=1em]{Figures/tree.png}RAPTOR &
34.82 & 23.20 & 52.52 & 5.33 &
29.51 & 29.87 & \cellcolor{LightGreen}\underline{29.50} & 4.33 \\
& \includegraphics[height=1em]{Figures/tree.png}GraphRAG &
35.10 & \cellcolor{LightGreen}\underline{26.10} & 56.89 & 3.00 &
32.73 & 54.25 & 26.48 & 4.33 \\

\specialrule{0.9pt}{0pt}{0pt}  

\multirow{6}{*}{\textbf{GraphSearch}} &
\includegraphics[height=1em]{Figures/doc.png}Dense &
\cellcolor{LightGreen}\underline{58.27} & 36.29 & 68.70 & 4.00 &
38.22 & 47.43 & 13.33 & 6.00 \\
& \includegraphics[height=1em]{Figures/hyper.png}HypergraphRAG &
51.04 & \cellcolor{LightGreen}\underline{44.72} & \cellcolor{LightGreen}\underline{69.97} & 3.33 &
\cellcolor{LightGreen}\underline{46.83} & \cellcolor{LightGreen}\underline{73.62} & 54.80 & 2.33 \\
& \includegraphics[height=1em]{Figures/graph.png}HippoRAG2 &
\cellcolor{LightBlue}\textbf{61.22} & 43.65 & \cellcolor{LightBlue}\textbf{72.47} & 1.67 &
\cellcolor{LightBlue}\textbf{58.64} & \cellcolor{LightBlue}\textbf{79.88} & \cellcolor{LightGreen}\underline{55.10} & 1.33 \\
& \includegraphics[height=1em]{Figures/linear.png}LinearRAG &
52.12 & \cellcolor{LightBlue}\textbf{44.77} & 68.52 & 4.00 &
41.65 & 70.26 & 49.35 & 4.33 \\
& \includegraphics[height=1em]{Figures/tree.png}RAPTOR &
53.80 & 42.38 & 68.56 & 4.33 &
40.14 & 71.24 & \cellcolor{LightBlue}\textbf{55.26} & 3.33 \\
& \includegraphics[height=1em]{Figures/tree.png}GraphRAG &
52.78 & 43.60 & 69.64 & 3.67 &
42.25 & 72.41 & 46.73 & 3.67 \\

\specialrule{1.2pt}{0pt}{0pt}  
\end{tabular}
\end{table*}

\subsubsection{Trained Search Agent.}
\textbf{\textit{\underline{Observation 3}}} \textit{RL-based training generally improves agentic performance across dense RAG and GraphRAG backends, but does not consistently outperform strong training-free agentic pipelines.} As shown in Table~\ref{tab:1}, RL-based agents (Search-R1 and Graph-R1) consistently improve over their corresponding training-free baselines on both general and multi-hop QA. However, these gains do not consistently surpass strong training-free pipelines with more structured workflows. In particular, GraphSearch-based systems—despite being training-free—generally outperform both Search-R1 and Graph-R1 variants by leveraging explicit design choices such as query decomposition and structured retrieval. These results indicate that while RL-based optimization can refine agentic behavior, well-designed training-free agentic workflows remain highly competitive and can even exceed RL-based systems.



To summarize, across multi-hop QA tasks, agentic search introduces implicit structural cues through iterative retrieval and reasoning, partially mitigating the absence of explicit graph structure in dense RAG. However, consistent with Observations~1–3, GraphRAG methods continue to deliver the strongest and most stable performance on complex multi-hop reasoning, indicating that explicit structural representations remain beneficial in this regime.
In contrast, for general QA, the performance gains from agentic search and GraphRAG are comparatively modest (e.g., average improvements of +2.43 vs.\ +26.25 for multi-hop QA). Given the substantial offline construction and latency overhead associated with GraphRAG (Table~\ref{tab:cost}), dense RAG—especially when paired with well-designed agentic workflows—remains a practical and competitive alternative for general QA scenarios.

\subsection{Training-free Agentic Workflow (RQ2)}
To answer \textbf{RQ2}, we examine how different retrieval backends affect performance under training-free agentic workflows. For each agentic method, we vary only the retrieval backend while keeping the agent design and inference protocol fixed. The results are reported in Table~\ref{tab:2}. We observe that \textbf{\textit{\underline{Observation 4}}} \textit{In training-free agentic workflows, explicit graph structure continues to deliver consistent and substantial benefits for multi-hop QA, while dense RAG remains competitive for general QA.} For multi-hop QA, GraphRAG-based backends consistently achieve the best or second-best results across both Search-o1 and GraphSearch, indicating that explicit structural representations remain effective for multi-hop reasoning even under agentic search. Among these methods, the entity-centric HippoRAG2 yields the largest gains over dense RAG. In contrast, for general QA, dense RAG remains highly competitive and in some cases outperforms certain GraphRAG variants (e.g., LinearRAG and RAPTOR on TriviaQA). This suggests that while graph structure is particularly beneficial for multi-hop reasoning, dense RAG continues to be a strong and practical choice for general QA within training-free agentic workflows.


\subsection{RL-based Search Agent (RQ3)}
\begin{figure}[ht!]
\centering
\includegraphics[width=7cm, height=5cm]{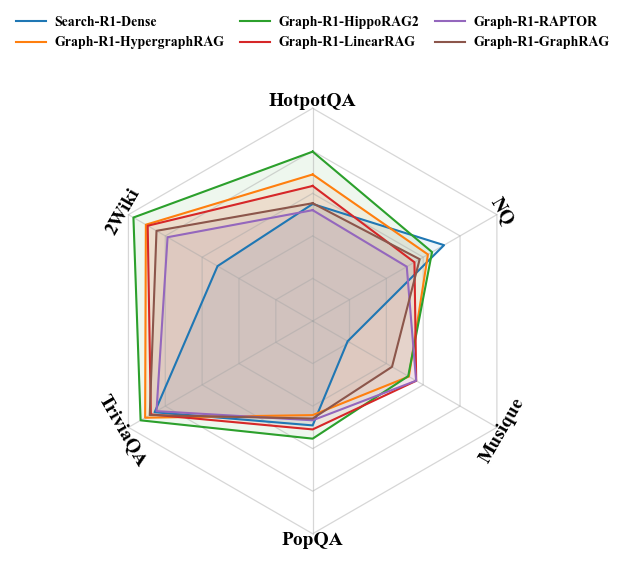}
\caption{Performance of different retrieval infrastructures on RL-based agentic systems}
\label{fig:3}
\end{figure}

We evaluate how policy learning via reinforcement learning interacts with different retrieval infrastructures across general and multi-hop QA settings. From Figure~\ref{fig:3}, we observe that: \textbf{\textit{\underline{Observation 5}}} \textit{RL-based agentic performance is highly backend-dependent: graph-based retrievers yield larger gains on multi-hop QA.} On multi-hop QA, GraphRAG-style backends consistently outperform dense RAG, though performance varies across graph-based methods. In particular, the entity-centric HippoRAG2 achieves the strongest results on datasets such as HotpotQA, PopQA, and 2Wiki, indicating that entity-level graph signals are especially effective for RL to exploit. In contrast, on general QA, dense RAG performs comparably to—or better than—several graph-based variants, notably achieving the best results on NQ. This suggests that even under RL-based agentic systems, dense RAG remains a strong baseline for general QA, while graph structure is most beneficial for multi-hop reasoning.

\begin{table*}[!h]
\caption{\textbf{Sensitivity Analysis on agentic systems}}
\label{tab:sensitivity}
\begin{tabular}{c|ccc|ccc}
\hline
\multirow{2}{*}{\textbf{Method}} &
\multicolumn{3}{c|}{\textbf{HotpotQA}} &
\multicolumn{3}{c}{\textbf{PopQA}} \\
\cline{2-7}
& \textbf{Search Turn} & \textbf{Recall} & \textbf{Variance}
& \textbf{Search Turn} & \textbf{Recall} & \textbf{Variance} \\
\hline
\textbf{Search-o1-Dense}     & 2.20 & 79.38 & 33.65$\pm$1.03 & 1.53 & 76.33 & 25.62$\pm$0.61 \\
\textbf{Search-o1-HippoRAG2} & 2.03 & 80.27 & 42.36$\pm$0.22 & 1.52 & 78.12 & 27.81$\pm$0.36 \\
\textbf{Search-R1}           & 1.82 & 81.67 & 34.82$\pm$0.95 & 1.36 & 77.15 & 33.15$\pm$0.54 \\
\textbf{Graph-R1-HippoRAG2}  & 1.71 & 83.50  & 53.71$\pm$0.18 & 1.38 & 78.61 & 36.13$\pm$0.32 \\
\hline
\end{tabular}
\end{table*}

\begin{figure*}[htbp]
    \centering

    \begin{subfigure}[b]{0.32\textwidth}
        \centering
        \includegraphics[width=\textwidth]{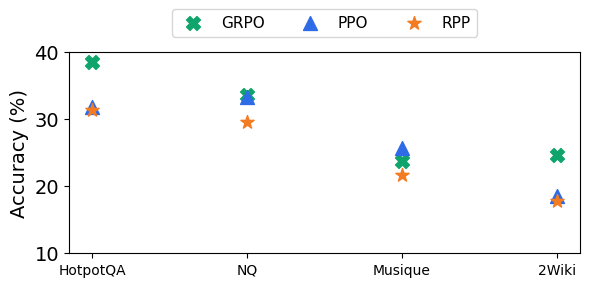}
        \caption{Graph-R1-HypergraphRAG-3B}
        \label{fig:a}
    \end{subfigure}\hfill
    \begin{subfigure}[b]{0.32\textwidth}
        \centering
        \includegraphics[width=\textwidth]{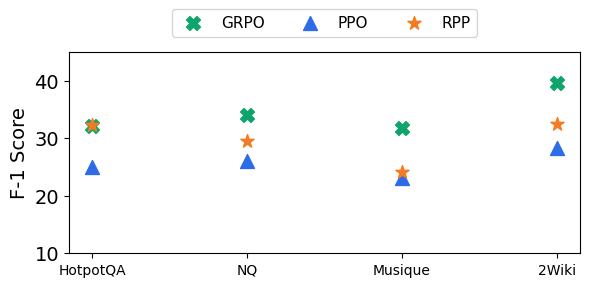}
        \caption{Graph-R1-LinearRAG-3B}
        \label{fig:b}
    \end{subfigure}\hfill
    \begin{subfigure}[b]{0.32\textwidth}
        \centering
        \includegraphics[width=\textwidth]{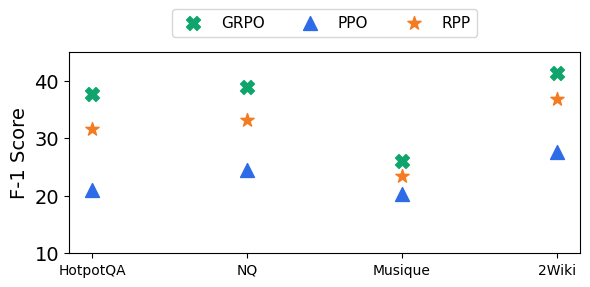}
        \caption{Graph-R1-HippoRAG2-3B}
        \label{fig:c}
    \end{subfigure}

    \caption{The impact of different RL algorithms}
    \label{fig:RL_algo}
\end{figure*}


\subsection{Sensitivity Analysis}
\subsubsection{Robust analysis} To answer \textbf{RQ4}, we investigate the robustness of dense RAG and GraphRAG under agentic search using retrieval recall and the mean and variance of Contain-EM (see Table~\ref{tab:sensitivity}).
\textbf{\textit{\underline{Observation 6}}} \textit{GraphRAG is more robust and stable than dense RAG in agentic search}.
Despite comparable search depths, GraphRAG achieves higher document hit rates and lower variance in Contain-EM, indicating more reliable evidence retrieval and greater stability in multi-turn agentic search scenarios.

\subsubsection{Impact of RL Training Paradigms}
To answer \textbf{RQ5}, we study how different RL training paradigms affect agentic search performance. We compare GRPO with alternative RL algorithms under identical agentic settings and retrieval backends. \textbf{\textit{\underline{Observation 7}}} \textit{GRPO is a favorable training paradigm for RL-based agentic systems}. As shown in Figure~\ref{fig:RL_algo}, GRPO consistently outperforms the other RL algorithms across both dense RAG and GraphRAG backends. This indicates that GRPO provides more effective policy optimization for agentic search, enabling agents to better exploit structured evidence and multi-step reasoning.

\subsubsection{Impact of Different scale of LLM Backbones}
We further analyze the influence of LLM backbone size (\textbf{RQ5}). In detail, for training-free workflows, we compare Qwen2.5-7B-Instruct and Qwen2.5-32B-Instruct. For RL-based paradigm, we implement a 3B variant with Qwen2.5-3B-Instruct. The results are shown in Table~\ref{tab:rl_backbone} and Table~\ref{tab:train_free_backbone}. \textbf{\textit{\underline{Observation 8}}} \textit{Larger backbones not only improve reasoning performance but also reduce the performance gap between GraphRAG and dense RAG.} In RL-based systems, scaling from 3B to 7B reduces the average GraphRAG–Dense gap from 14.70 to 9.75 (e.g., HotpotQA 19.08→15.99, PopQA 5.74→3.05). A similar trend appears in training-free systems, where increasing the backbone from 7B to 32B slightly decreases the average gap from 7.80 to 7.19. These results suggest that stronger LLMs can better leverage implicit structural cues through reasoning, partially compensating for the absence of explicit graph structure.

\begin{table}[!h]
\centering
\small
\caption{3B vs 7B on RL-based agentic systems.}
\begin{tabular}{lcccc}
\specialrule{1.2pt}{0pt}{0pt}  
Method & HotpotQA & Musique & NQ & PopQA \\
\midrule
Search-R1-3B & 23.67 & 4.96 & 30.50 & 25.34 \\
Graph-R1-3B$^\spadesuit$ & 42.75 & 31.73 & 37.72 & 31.08 \\
Search-R1-7B & 35.76 & 12.35 & 46.35 & 32.90 \\
Graph-R1-7B$^\spadesuit$ & 51.75 & 36.53 & 42.12 & 35.95 \\
\specialrule{1.2pt}{0pt}{0pt}  
\end{tabular}
\vspace{-10pt}
\label{tab:rl_backbone}
\end{table}

\begin{table}[!h]
\centering
\small
\caption{7B vs 32B on training-free systems.}
\begin{tabular}{lcccc}
\specialrule{1.2pt}{0pt}{0pt}  
Method & HotpotQA & Musique & NQ & PopQA \\
\midrule
Search-o1-Dense-7B & 33.76 & 12.62 & 38.20 & 25.78 \\
Search-o1-GraphRAG-7B$^\spadesuit$ & 42.75 & 32.44 & 38.34 & 28.01 \\
Search-o1-Dense-32B & 40.85 & 18.95 & 51.80 & 36.20 \\
Search-o1-GraphRAG-32B$^\spadesuit$ & 47.01 & 42.37 & 50.50 & 36.69 \\
\specialrule{1.2pt}{0pt}{0pt}  
\end{tabular}
\label{tab:train_free_backbone}
\end{table}

\section{Conclusions}
This work introduces RAGSearch, a benchmark for systematically evaluating dense RAG and representative GraphRAG pipelines as retrieval infrastructures within agentic search systems. Our results show that while agentic search can partially compensate for missing structure in dense RAG through iterative retrieval and reasoning, explicit graph-based retrieval remains crucial for robust multi-hop reasoning. GraphRAG methods consistently deliver stronger performance and greater stability in complex settings, whereas dense RAG remains a practical and competitive choice for general QA due to its lower construction cost.
More broadly, our findings suggest that agentic reasoning is reshaping the role of retrieval structure in LLM-based systems. Rather than replacing explicit structure, agentic search redistributes where structure emerges—shifting some from offline graph construction to online interaction. Understanding this balance between explicit and implicit structure will be key to designing the next generation of retrieval-augmented and agentic AI systems, and we hope RAGSearch will facilitate further research on this emerging design space.
\newpage
\bibliographystyle{ACM-Reference-Format}
\bibliography{reference}

@article{yu2025graphragr1,
  title={Graphrag-r1: Graph retrieval-augmented generation with process-constrained reinforcement learning},
  author={Yu, Chuanyue and Zhao, Kuo and Li, Yuhan and Chang, Heng and Feng, Mingjian and Jiang, Xiangzhe and Sun, Yufei and Li, Jia and Zhang, Yuzhi and Li, Jianxin and others},
  journal={arXiv preprint arXiv:2507.23581},
  year={2025}
}

@article{jin2025searchr1,
  title={Search-r1: Training llms to reason and leverage search engines with reinforcement learning},
  author={Jin, Bowen and Zeng, Hansi and Yue, Zhenrui and Yoon, Jinsung and Arik, Sercan and Wang, Dong and Zamani, Hamed and Han, Jiawei},
  journal={arXiv preprint arXiv:2503.09516},
  year={2025}
}

@article{dong2025youtu,
  title={Youtu-graphrag: Vertically unified agents for graph retrieval-augmented complex reasoning},
  author={Dong, Junnan and An, Siyu and Yu, Yifei and Zhang, Qian-Wen and Luo, Linhao and Huang, Xiao and Wu, Yunsheng and Yin, Di and Sun, Xing},
  journal={arXiv preprint arXiv:2508.19855},
  year={2025}
}

@inproceedings{lee2025hybgrag,
  title={Hybgrag: Hybrid retrieval-augmented generation on textual and relational knowledge bases},
  author={Lee, Meng-Chieh and Zhu, Qi and Mavromatis, Costas and Han, Zhen and Adeshina, Soji and Ioannidis, Vassilis N and Rangwala, Huzefa and Faloutsos, Christos},
  booktitle={Proceedings of the 63rd Annual Meeting of the Association for Computational Linguistics (Volume 1: Long Papers)},
  pages={879--893},
  year={2025}
}

@article{li2025searcho1,
  title={Search-o1: Agentic search-enhanced large reasoning models},
  author={Li, Xiaoxi and Dong, Guanting and Jin, Jiajie and Zhang, Yuyao and Zhou, Yujia and Zhu, Yutao and Zhang, Peitian and Dou, Zhicheng},
  journal={arXiv preprint arXiv:2501.05366},
  year={2025}
}

@article{yang2025graphsearch,
  title={GraphSearch: An Agentic Deep Searching Workflow for Graph Retrieval-Augmented Generation},
  author={Yang, Cehao and Wu, Xiaojun and Lin, Xueyuan and Xu, Chengjin and Jiang, Xuhui and Sun, Yuanliang and Li, Jia and Xiong, Hui and Guo, Jian},
  journal={arXiv preprint arXiv:2509.22009},
  year={2025}
}

@inproceedings{yao2022react,
  title={React: Synergizing reasoning and acting in language models},
  author={Yao, Shunyu and Zhao, Jeffrey and Yu, Dian and Du, Nan and Shafran, Izhak and Narasimhan, Karthik R and Cao, Yuan},
  booktitle={The eleventh international conference on learning representations},
  year={2022}
}

@article{luo2025graphr1,
  title={Graph-r1: Towards agentic graphrag framework via end-to-end reinforcement learning},
  author={Luo, Haoran and Chen, Guanting and Lin, Qika and Guo, Yikai and Xu, Fangzhi and Kuang, Zemin and Song, Meina and Wu, Xiaobao and Zhu, Yifan and Tuan, Luu Anh and others},
  journal={arXiv preprint arXiv:2507.21892},
  year={2025}
}

@article{han2024grapgrag,
  title={Retrieval-augmented generation with graphs (graphrag)},
  author={Han, Haoyu and Wang, Yu and Shomer, Harry and Guo, Kai and Ding, Jiayuan and Lei, Yongjia and Halappanavar, Mahantesh and Rossi, Ryan A and Mukherjee, Subhabrata and Tang, Xianfeng and others},
  journal={arXiv preprint arXiv:2501.00309},
  year={2024}
}

@article{edge2024local,
  title={From local to global: A graph rag approach to query-focused summarization},
  author={Edge, Darren and Trinh, Ha and Cheng, Newman and Bradley, Joshua and Chao, Alex and Mody, Apurva and Truitt, Steven and Metropolitansky, Dasha and Ness, Robert Osazuwa and Larson, Jonathan},
  journal={arXiv preprint arXiv:2404.16130},
  year={2024}
}

@misc{lewis2021rag,
      title={Retrieval-Augmented Generation for Knowledge-Intensive NLP Tasks}, 
      author={Patrick Lewis and Ethan Perez and Aleksandra Piktus and Fabio Petroni and Vladimir Karpukhin and Naman Goyal and Heinrich Küttler and Mike Lewis and Wen-tau Yih and Tim Rocktäschel and Sebastian Riedel and Douwe Kiela},
      year={2021},
      eprint={2005.11401},
      archivePrefix={arXiv},
      primaryClass={cs.CL},
      url={https://arxiv.org/abs/2005.11401}, 
}

@misc{jeong2024adaptiverag,
      title={Adaptive-RAG: Learning to Adapt Retrieval-Augmented Large Language Models through Question Complexity}, 
      author={Soyeong Jeong and Jinheon Baek and Sukmin Cho and Sung Ju Hwang and Jong C. Park},
      year={2024},
      eprint={2403.14403},
      archivePrefix={arXiv},
      primaryClass={cs.CL},
      url={https://arxiv.org/abs/2403.14403}, 
}

@misc{wang2024maferw,
      title={MaFeRw: Query Rewriting with Multi-Aspect Feedbacks for Retrieval-Augmented Large Language Models}, 
      author={Yujing Wang and Hainan Zhang and Liang Pang and Binghui Guo and Hongwei Zheng and Zhiming Zheng},
      year={2024},
      eprint={2408.17072},
      archivePrefix={arXiv},
      primaryClass={cs.CL},
      url={https://arxiv.org/abs/2408.17072}, 
}

@misc{sarthi2024raptor,
      title={RAPTOR: Recursive Abstractive Processing for Tree-Organized Retrieval}, 
      author={Parth Sarthi and Salman Abdullah and Aditi Tuli and Shubh Khanna and Anna Goldie and Christopher D. Manning},
      year={2024},
      eprint={2401.18059},
      archivePrefix={arXiv},
      primaryClass={cs.CL},
      url={https://arxiv.org/abs/2401.18059}, 
}

@misc{gutiérrez2025hipporag,
      title={HippoRAG: Neurobiologically Inspired Long-Term Memory for Large Language Models}, 
      author={Bernal Jiménez Gutiérrez and Yiheng Shu and Yu Gu and Michihiro Yasunaga and Yu Su},
      year={2025},
      eprint={2405.14831},
      archivePrefix={arXiv},
      primaryClass={cs.CL},
      url={https://arxiv.org/abs/2405.14831}, 
}

@misc{gutiérrez2025hipporag2,
      title={From RAG to Memory: Non-Parametric Continual Learning for Large Language Models}, 
      author={Bernal Jiménez Gutiérrez and Yiheng Shu and Weijian Qi and Sizhe Zhou and Yu Su},
      year={2025},
      eprint={2502.14802},
      archivePrefix={arXiv},
      primaryClass={cs.CL},
      url={https://arxiv.org/abs/2502.14802}, 
}

@misc{guo2025lightrag,
      title={LightRAG: Simple and Fast Retrieval-Augmented Generation}, 
      author={Zirui Guo and Lianghao Xia and Yanhua Yu and Tu Ao and Chao Huang},
      year={2025},
      eprint={2410.05779},
      archivePrefix={arXiv},
      primaryClass={cs.IR},
      url={https://arxiv.org/abs/2410.05779}, 
}

@misc{he2024gretriever,
      title={G-Retriever: Retrieval-Augmented Generation for Textual Graph Understanding and Question Answering}, 
      author={Xiaoxin He and Yijun Tian and Yifei Sun and Nitesh V. Chawla and Thomas Laurent and Yann LeCun and Xavier Bresson and Bryan Hooi},
      year={2024},
      eprint={2402.07630},
      archivePrefix={arXiv},
      primaryClass={cs.LG},
      url={https://arxiv.org/abs/2402.07630}, 
}

@misc{luo2025hypergraphrag,
      title={HyperGraphRAG: Retrieval-Augmented Generation via Hypergraph-Structured Knowledge Representation}, 
      author={Haoran Luo and Haihong E and Guanting Chen and Yandan Zheng and Xiaobao Wu and Yikai Guo and Qika Lin and Yu Feng and Zemin Kuang and Meina Song and Yifan Zhu and Luu Anh Tuan},
      year={2025},
      eprint={2503.21322},
      archivePrefix={arXiv},
      primaryClass={cs.AI},
      url={https://arxiv.org/abs/2503.21322}, 
}

@misc{zhuang2025linearrag,
      title={LinearRAG: Linear Graph Retrieval Augmented Generation on Large-scale Corpora}, 
      author={Luyao Zhuang and Shengyuan Chen and Yilin Xiao and Huachi Zhou and Yujing Zhang and Hao Chen and Qinggang Zhang and Xiao Huang},
      year={2025},
      eprint={2510.10114},
      archivePrefix={arXiv},
      primaryClass={cs.CL},
      url={https://arxiv.org/abs/2510.10114}, 
}

@misc{openai2024openaio1,
      title={OpenAI o1 System Card}, 
      author={OpenAI and : and Aaron Jaech and Adam Kalai and Adam Lerer and Adam Richardson and Ahmed El-Kishky and Aiden Low and Alec Helyar and Aleksander Madry and Alex Beutel and Alex Carney and Alex Iftimie and Alex Karpenko et al.},
      year={2024},
      eprint={2412.16720},
      archivePrefix={arXiv},
      primaryClass={cs.AI},
      url={https://arxiv.org/abs/2412.16720}, 
}

@misc{feng2025hyperrag,
      title={Hyper-RAG: Combating LLM Hallucinations using Hypergraph-Driven Retrieval-Augmented Generation}, 
      author={Yifan Feng and Hao Hu and Xingliang Hou and Shiquan Liu and Shihui Ying and Shaoyi Du and Han Hu and Yue Gao},
      year={2025},
      eprint={2504.08758},
      archivePrefix={arXiv},
      primaryClass={cs.IR},
      url={https://arxiv.org/abs/2504.08758}, 
}

@misc{liu2026graphsearch,
      title={GraphSearch: Agentic Search-Augmented Reasoning for Zero-Shot Graph Learning}, 
      author={Jiajin Liu and Yuanfu Sun and Dongzhe Fan and Qiaoyu Tan},
      year={2026},
      eprint={2601.08621},
      archivePrefix={arXiv},
      primaryClass={cs.CL},
      url={https://arxiv.org/abs/2601.08621}, 
}

@misc{chen2025pathragpruninggraphbasedretrieval,
      title={PathRAG: Pruning Graph-based Retrieval Augmented Generation with Relational Paths}, 
      author={Boyu Chen and Zirui Guo and Zidan Yang and Yuluo Chen and Junze Chen and Zhenghao Liu and Chuan Shi and Cheng Yang},
      year={2025},
      eprint={2502.14902},
      archivePrefix={arXiv},
      primaryClass={cs.CL},
      url={https://arxiv.org/abs/2502.14902}, 
}

@misc{luo2025kbqao1agenticknowledgebase,
      title={KBQA-o1: Agentic Knowledge Base Question Answering with Monte Carlo Tree Search}, 
      author={Haoran Luo and Haihong E and Yikai Guo and Qika Lin and Xiaobao Wu and Xinyu Mu and Wenhao Liu and Meina Song and Yifan Zhu and Luu Anh Tuan},
      year={2025},
      eprint={2501.18922},
      archivePrefix={arXiv},
      primaryClass={cs.CL},
      url={https://arxiv.org/abs/2501.18922}, 
}

@misc{shao2024grpo,
      title={DeepSeekMath: Pushing the Limits of Mathematical Reasoning in Open Language Models}, 
      author={Zhihong Shao and Peiyi Wang and Qihao Zhu and Runxin Xu and Junxiao Song and Xiao Bi and Haowei Zhang and Mingchuan Zhang and Y. K. Li and Y. Wu and Daya Guo},
      year={2024},
      eprint={2402.03300},
      archivePrefix={arXiv},
      primaryClass={cs.CL},
      url={https://arxiv.org/abs/2402.03300}, 
}

@misc{wang2025kblam,
      title={KBLaM: Knowledge Base augmented Language Model}, 
      author={Xi Wang and Taketomo Isazawa and Liana Mikaelyan and James Hensman},
      year={2025},
      eprint={2410.10450},
      archivePrefix={arXiv},
      primaryClass={cs.AI},
      url={https://arxiv.org/abs/2410.10450}, 
}

@misc{islam2024openrag,
      title={Open-RAG: Enhanced Retrieval-Augmented Reasoning with Open-Source Large Language Models}, 
      author={Shayekh Bin Islam and Md Asib Rahman and K S M Tozammel Hossain and Enamul Hoque and Shafiq Joty and Md Rizwan Parvez},
      year={2024},
      eprint={2410.01782},
      archivePrefix={arXiv},
      primaryClass={cs.CL},
      url={https://arxiv.org/abs/2410.01782}, 
}

@misc{he2024retrievingrethinkingrevisingchainofverification,
      title={Retrieving, Rethinking and Revising: The Chain-of-Verification Can Improve Retrieval Augmented Generation}, 
      author={Bolei He and Nuo Chen and Xinran He and Lingyong Yan and Zhenkai Wei and Jinchang Luo and Zhen-Hua Ling},
      year={2024},
      eprint={2410.05801},
      archivePrefix={arXiv},
      primaryClass={cs.CL},
      url={https://arxiv.org/abs/2410.05801}, 
}

@misc{zhang2024raftadaptinglanguagemodel,
      title={RAFT: Adapting Language Model to Domain Specific RAG}, 
      author={Tianjun Zhang and Shishir G. Patil and Naman Jain and Sheng Shen and Matei Zaharia and Ion Stoica and Joseph E. Gonzalez},
      year={2024},
      eprint={2403.10131},
      archivePrefix={arXiv},
      primaryClass={cs.CL},
      url={https://arxiv.org/abs/2403.10131}, 
}

@misc{zhang2025arise,
      title={ARise: Towards Knowledge-Augmented Reasoning via Risk-Adaptive Search}, 
      author={Yize Zhang and Tianshu Wang and Sirui Chen and Kun Wang and Xingyu Zeng and Hongyu Lin and Xianpei Han and Le Sun and Chaochao Lu},
      year={2025},
      eprint={2504.10893},
      archivePrefix={arXiv},
      primaryClass={cs.AI},
      url={https://arxiv.org/abs/2504.10893}, 
}

@inproceedings{FlashRAG,
  author       = {Jiajie Jin and
                  Yutao Zhu and
                  Zhicheng Dou and
                  Guanting Dong and
                  Xinyu Yang and
                  Chenghao Zhang and
                  Tong Zhao and
                  Zhao Yang and
                  Ji{-}Rong Wen},
  editor       = {Guodong Long and
                  Michale Blumestein and
                  Yi Chang and
                  Liane Lewin{-}Eytan and
                  Zi Helen Huang and
                  Elad Yom{-}Tov},
  title        = {FlashRAG: {A} Modular Toolkit for Efficient Retrieval-Augmented Generation
                  Research},
  booktitle    = {Companion Proceedings of the {ACM} on Web Conference 2025, {WWW} 2025,
                  Sydney, NSW, Australia, 28 April 2025 - 2 May 2025},
  pages        = {737--740},
  publisher    = {{ACM}},
  year         = {2025},
  url          = {https://doi.org/10.1145/3701716.3715313},
  doi          = {10.1145/3701716.3715313}
}

@misc{2wiki,
      title={Constructing A Multi-hop QA Dataset for Comprehensive Evaluation of Reasoning Steps}, 
      author={Xanh Ho and Anh-Khoa Duong Nguyen and Saku Sugawara and Akiko Aizawa},
      year={2020},
      eprint={2011.01060},
      archivePrefix={arXiv},
      primaryClass={cs.CL},
      url={https://arxiv.org/abs/2011.01060}, 
}

@misc{hotpotqa,
      title={HotpotQA: A Dataset for Diverse, Explainable Multi-hop Question Answering}, 
      author={Zhilin Yang and Peng Qi and Saizheng Zhang and Yoshua Bengio and William W. Cohen and Ruslan Salakhutdinov and Christopher D. Manning},
      year={2018},
      eprint={1809.09600},
      archivePrefix={arXiv},
      primaryClass={cs.CL},
      url={https://arxiv.org/abs/1809.09600}, 
}

@misc{musique,
      title={MuSiQue: Multihop Questions via Single-hop Question Composition}, 
      author={Harsh Trivedi and Niranjan Balasubramanian and Tushar Khot and Ashish Sabharwal},
      year={2022},
      eprint={2108.00573},
      archivePrefix={arXiv},
      primaryClass={cs.CL},
      url={https://arxiv.org/abs/2108.00573}, 
}

@article{nq,
    title = "Natural Questions: A Benchmark for Question Answering Research",
    author = "Kwiatkowski, Tom  and
      Palomaki, Jennimaria  and
      Redfield, Olivia  and
      Collins, Michael  and
      Parikh, Ankur  and
      Alberti, Chris  and
      Epstein, Danielle  and
      Polosukhin, Illia  and
      Devlin, Jacob  and
      Lee, Kenton  and
      Toutanova, Kristina  and
      Jones, Llion  and
      Kelcey, Matthew  and
      Chang, Ming-Wei  and
      Dai, Andrew M.  and
      Uszkoreit, Jakob  and
      Le, Quoc  and
      Petrov, Slav",
    editor = "Lee, Lillian  and
      Johnson, Mark  and
      Roark, Brian  and
      Nenkova, Ani",
    journal = "Transactions of the Association for Computational Linguistics",
    volume = "7",
    year = "2019",
    address = "Cambridge, MA",
    publisher = "MIT Press",
    url = "https://aclanthology.org/Q19-1026/",
    doi = "10.1162/tacl_a_00276",
    pages = "452--466"
}

@misc{popqa,
      title={When Not to Trust Language Models: Investigating Effectiveness of Parametric and Non-Parametric Memories}, 
      author={Alex Mallen and Akari Asai and Victor Zhong and Rajarshi Das and Daniel Khashabi and Hannaneh Hajishirzi},
      year={2023},
      eprint={2212.10511},
      archivePrefix={arXiv},
      primaryClass={cs.CL},
      url={https://arxiv.org/abs/2212.10511}, 
}

@misc{triviaqa,
      title={TriviaQA: A Large Scale Distantly Supervised Challenge Dataset for Reading Comprehension}, 
      author={Mandar Joshi and Eunsol Choi and Daniel S. Weld and Luke Zettlemoyer},
      year={2017},
      eprint={1705.03551},
      archivePrefix={arXiv},
      primaryClass={cs.CL},
      url={https://arxiv.org/abs/1705.03551}, 
}

@misc{wikidump,
      title={Dense Passage Retrieval for Open-Domain Question Answering}, 
      author={Vladimir Karpukhin and Barlas Oğuz and Sewon Min and Patrick Lewis and Ledell Wu and Sergey Edunov and Danqi Chen and Wen-tau Yih},
      year={2020},
      eprint={2004.04906},
      archivePrefix={arXiv},
      primaryClass={cs.CL},
      url={https://arxiv.org/abs/2004.04906}, 
}

@misc{qwen2025qwen25technicalreport,
      title={Qwen2.5 Technical Report}, 
      author={Qwen and : and An Yang and Baosong Yang and Beichen Zhang and Binyuan Hui and Bo Zheng and Bowen Yu and Chengyuan Li and Dayiheng Liu and Fei Huang and Haoran Wei, et al.},
      year={2025},
      eprint={2412.15115},
      archivePrefix={arXiv},
      primaryClass={cs.CL},
      url={https://arxiv.org/abs/2412.15115}, 
}

\appendix
\section{Datasets Statistics}
\label{app:a}
We conduct evaluations on six widely used RAG benchmarks from the FlashRAG toolkit~\cite{FlashRAG}, covering both single-hop and multi-hop question answering tasks:
\begin{itemize}
    \item \textbf{Natural Questions (NQ)~\cite{nq}.} Real user questions from Google Search paired with Wikipedia passages/answers; commonly used for open-domain, mostly single-hop QA.
    \item \textbf{PopQA~\cite{popqa}.} Popular-knowledge question set designed for retrieval-based QA, emphasizing factual queries where the answer must be grounded in retrieved evidence.
    \item \textbf{TriviaQA~\cite{triviaqa}.} Trivia-style questions with evidence documents (often web/Wikipedia); used for open-domain factual QA and long-context evidence matching.
    \item \textbf{HotpotQA~\cite{hotpotqa}.} Multi-hop QA requiring reasoning over multiple supporting Wikipedia passages; includes labeled supporting facts.
    \item \textbf{Musique~\cite{musique}} Multi-hop QA benchmark built to test compositional reasoning across several pieces of evidence, often with more challenging, structured multi-step requirements.
    \item \textbf{2WikiMultiHopQA (2Wiki)~\cite{2wiki}} Multi-hop QA constructed from Wikipedia that typically requires linking two (or more) pages to reach the answer, focusing on cross-article reasoning.
\end{itemize}
The detailed dataset statistics are in Table~\ref{tab:dataset}. For RL-based agentic systems, we randomly sample 5000 data from the Train set of HotpotQA and NQ. For the test set of RAGSearch, we adopt the full original Dev or Test set. In detail, for NQ, PopQA and TriviaQA, we choose the Test set; For HotpotQA, Musique, 2Wiki, we choose the Dev set.
\begin{table*}[!h]
\caption{Dataset Statistics}
\label{tab:dataset}
\begin{tabular}{c|c|c|ccc}
\hline
Dataset  & Task         & Knowledge Source & \#Train                                               & \#Dev                                                & \#Test                                           \\ \hline
NQ       & General QA   & Wiki             & 79,168                                                & 8,757                                                & 3,610                                            \\
PopQA    & General QA   & Wiki             & -                                                     & -                                                    & 14,267                                           \\
TriviaQA & General QA   & Wiki \& Web   & 78,785   & 8,837 & 11,313                                           \\
HotpotQA & Multi-hop QA & Wiki             & 90,447                                                & 7,405                                                & -                                                \\
Musique  & Multi-hop QA & Wiki             & 19,938                                                &  2,417 &  - \\
2Wiki    & Multi-hop QA & Wiki             & 15,000                                                & 12,576                                               & -                                                \\ \hline
\end{tabular}
\end{table*}
\section{Implementation Details}
\label{app:b}
To conduct a comprehensive evaluation, we include two different categories of agentic systems for assessment. Here is brief introduction to all the method involved.
\paragraph{\textbf{Training-free Workflows:}}
These approaches do not train an explicit control policy; rather, they use structured prompts and heuristic rules to steer multi-step retrieval and reasoning at inference time. Specifically, we evaluate;
\begin{itemize}
    \item Search-o1~\cite{li2025searcho1}: Augments large reasoning models with an agentic RAG workflow and a Reason-in-Documents module that refines retrieved evidence before integration, enabling dynamic, noise-reduced knowledge retrieval to improve reliability on complex reasoning tasks and open-domain QA. Our implementation is based on \url{https://github.com/RUC-NLPIR/Search-o1}
    \item GraphSearch~\cite{yang2025graphsearch}: An agentic deep-search workflow for GraphRAG that performs multi-turn, modular retrieval with dual-channel querying over both text chunks (semantic) and structural graphs (relational), consistently improving multi-hop RAG accuracy and generation quality over traditional GraphRAG retrieval. Our implementation is based on \url{https://github.com/DataArcTech/GraphSearch}
\end{itemize}
\paragraph{\textbf{Trained Search Agent:}}
These approaches use an RL policy to optimize the agent’s search and reasoning behavior, typically adopting GRPO as the reinforcement learning algorithm. Specifically, we choose:
\begin{itemize}
    \item Search-R1~\cite{jin2025searchr1}: A RL-based retrieval-augmented reasoning framework that trains LLMs to autonomously generate multi-turn search queries during step-by-step reasoning, using retrieved-token masking and an outcome-based reward to improve QA performance over standard RAG baselines. Our implementation is based on \url{https://github.com/PeterGriffinJin/Search-R1}
    \item Graph-R1~\cite{luo2025graphr1}: An agentic GraphRAG framework trained end-to-end with reinforcement learning that builds lightweight knowledge hypergraphs and performs multi-turn retrieval as an agent–environment interaction. Our implementation is based on \url{https://github.com/LHRLAB/Graph-R1}
\end{itemize}
For all the RL-based methods, we pre-train the model with GRPO for 3 epochs. We set the batch size as 32, max search turn as 5.

For retrieve backends, we mainly adopt 5 representative GraphRAGs:
\begin{itemize}
    \item \textbf{HypergraphRAG}~\cite{luo2025hypergraphrag}: A hypergraph-based RAG framework that represents real-world n-ary facts using hyperedges and integrates hypergraph construction, retrieval, and generation. Our implementation is based on \url{https://github.com/LHRLAB/Graph-R1}
    \item \textbf{HippoRAG2}~\cite{gutiérrez2025hipporag2}:A memory-inspired RAG framework that extends HippoRAG’s Personalized PageRank retrieval with deeper passage integration and stronger online LLM usage, improving factual, sense-making, and associative memory. Our implementation is based on \url{https://github.com/OSU-NLP-Group/HippoRAG}
    \item \textbf{LinearRAG}~\cite{zhuang2025linearrag}: An efficient GraphRAG framework that avoids noisy, costly relation extraction by building a lightweight relation-free hierarchical “Tri-Graph” (via entity extraction + semantic linking) and retrieving evidence with a two-stage process—local entity activation followed by global importance aggregation—yielding stronger and more reliable passage retrieval on multi-hop QA benchmarks. Our implementation is based on \url{https://github.com/DEEP-PolyU/LinearRAG}
    \item \textbf{RAPTOR}~\cite{sarthi2024raptor}: A retrieval-augmented approach that builds a hierarchical tree of recursive embeddings, clusters, and bottom-up summaries, enabling inference-time retrieval across long documents at multiple abstraction levels and delivering strong gains. Our implementation is based on \url{https://github.com/parthsarthi03/raptor}
    \item \textbf{GraphRAG}~\cite{edge2024local}:A graph-based QA framework for private corpora that tackles global, corpus-level questions by (1) building an entity knowledge graph and precomputing community summaries, then (2) answering queries via summary-to-partial-response generation followed by a final aggregation, improving comprehensiveness and diversity over standard RAG at million-token scale. ur implementation is based on \url{https://microsoft.github.io/graphrag/}
\end{itemize}
For all the GraphRAG, we utilize the context of each question as the document and organize the corpus by the official settings. For retrieval, we set the top-k as top-5.
\section{Details of Evaluation Metrics}
\label{app:eval_metric}
\noindent\textbf{Contain Exact Match}~\cite{zhuang2025linearrag}: Check if the correct answer appears in the generated response
\[
\mathrm{Contain\ Exact\ Match}=\frac{1}{N}\sum_{i=1}^{N}\mathbf{1}\big(\mathrm{norm}(a_i)\subseteq \mathrm{norm}(\hat{y}_i)\big)
\]
\noindent\textbf{F-1}: The F1 score evaluates the token-level overlap between the predicted answer $y_i$ and the ground-truth answer $y_i^\star$ using the harmonic mean of precision and recall
\[
\mathrm{F1}=\frac{1}{N}\sum_{i=1}^{N}
\frac{2\cdot\left|\mathrm{tokens}(y_i)\cap \mathrm{tokens}(y_i^\star)\right|}
{\left|\mathrm{tokens}(y_i)\right|+\left|\mathrm{tokens}(y_i^\star)\right|}
\]
\section{F-1 Score for Agentic Search Systems}
In this section, we report the F-1 score of different agentic systems. As Table~\ref{tab:f-1} indicates, both agentic systems can narrow the performance gap between GraphRAG and dense RAG in multi-hop QA settings. For general QA settings, dense-RAG is still a competitive retrieve backend.
\begin{table*}[t]
\centering
\small
\setlength{\tabcolsep}{2pt}
\renewcommand{\arraystretch}{1.10}
\caption{Overall F-1 of  different systems}
\label{tab:f-1}

\begin{tabular}{@{} L{2.2cm} L{3.8cm} *{4}{C{1.5cm}} @{}}
\toprule
\multirow{2}{*}{\textbf{System}} &
\multirow{2}{*}{\textbf{Methods}} &
\multicolumn{2}{c}{\textbf{General QA}} &
\multicolumn{2}{c}{\textbf{Multi-Hop QA}} \\
\cmidrule(r){3-4}\cmidrule(l){5-6}
& & \textbf{NQ$^\dagger$} & \textbf{TriviaQA$^\star$}
  & \textbf{HotpotQA$^\dagger$} & \textbf{Musique$^\star$} \\
\midrule

\multirow{2}{*}{\textbf{Single-shot}} &
Qwen-2.5-7B-Dense & 39.3 & 61.12 & 20.12 & 29.08 \\
& Qwen-2.5-7B-GraphRAG$^\spadesuit$ &
27.92\downdiff{11.38} & 49.25\downdiff{11.87} &
33.72\updiff{13.60} & 41.13\updiff{12.05} \\
\midrule

\multirow{4}{*}{\textbf{Training-free}} &
Search-o1-7B-Dense & 36.53 & 59.73 & 39.08 & 17.46 \\
& Search-o1-7B-GraphRAG$^\spadesuit$ &
36.62\updiff{0.09} & 58.18\downdiff{1.55} &
41.57\updiff{2.49} & 37.01\updiff{19.55} \\
& GraphSearch-7B-Dense & 8.70 & 13.12 & 6.02 & 2.36 \\
& GraphSearch-7B-GraphRAG$^\spadesuit$ &
4.61\downdiff{4.09} & 14.18\updiff{1.06} &
5.48\downdiff{0.54} & 4.21\updiff{1.85} \\
\midrule

\multirow{2}{*}{\textbf{RL-based}} &
Search-R1-7B & 47.26 & 59.21 & 37.13 & 16.02 \\
& Graph-R1-7B$^\spadesuit$ &
44.21\downdiff{3.05} & 62.10\updiff{2.89} &
43.25\updiff{5.12} & 35.12\updiff{19.10} \\
\bottomrule
\end{tabular}
\end{table*}

\section{Efficiency of GraphRAGs}
We report the knowledge construction cost and offline inference time of different graph-based retriever. 
\begin{table*}[!h]
\caption{Comparison on cost of GraphRAGs of NQ. TM indicates construction time per 1M token; CM is the cost per 1M token, RT is the average retrieval time and CT is the average context length.}
\label{tab:cost}
\begin{tabular}{c|cccc}
\hline
\multirow{2}{*}{Method} & \multicolumn{2}{c}{Knowledge Construction} & \multicolumn{2}{c}{Offline Inference} \\ \cline{2-5} 
                        & TM       & \multicolumn{1}{c|}{CM}         & \multicolumn{1}{c|}{RT/s}  & CT/token \\ \hline
HypergraphRAG           & 1.37h    & \multicolumn{1}{c|}{3.93\$}     & \multicolumn{1}{c|}{0.77}  & 1680     \\
HippoRAG2               & 1.19h    & \multicolumn{1}{c|}{2.85\$}     & \multicolumn{1}{c|}{1.00}     & 3229     \\
LinearRAG               & 0.68h    & \multicolumn{1}{c|}{0\$}        & \multicolumn{1}{c|}{1.18}  & 4600     \\
RAPTOR                  & 1.70h    & \multicolumn{1}{c|}{6.38\$}     & \multicolumn{1}{c|}{8.4}   & 814      \\
GraphRAG                & 1.72h    & \multicolumn{1}{c|}{13.19\$}    & \multicolumn{1}{c|}{1.16}  & 22160    \\ \hline
\end{tabular}
\end{table*}
\section{Case Study}
In this section, we displays the difference between different GraphRAGs.

\begin{table*}[htbp]
\centering
\caption{Case study of different agentic systems.}
\label{tab:case}
\setlength{\tabcolsep}{10pt}       
\renewcommand{\arraystretch}{1.35} 

{\scriptsize 
\begin{tabular}{|>{\centering\arraybackslash}m{1.5cm}|
                    >{\centering\arraybackslash}m{4cm}|
                    >{\centering\arraybackslash}m{4cm}|
                    >{\centering\arraybackslash}m{4cm}|}
\hline

\noalign{\renewcommand{\arraystretch}{0.675}}
\cellcolor{gray!20}\textbf{Question} &
\multicolumn{3}{>{\columncolor{gray!20}\centering\arraybackslash}m{\dimexpr 12cm + 4\tabcolsep\relax}|}
{\textbf{"Are the Laleli Mosque and Esma Sultan Mansion located in the same neighborhood?"}}\\
\hline

\cellcolor{gray!20}\textbf{Golden Answer} &
\multicolumn{3}{>{\columncolor{gray!20}\centering\arraybackslash}m{\dimexpr 12cm + 4\tabcolsep\relax}|}
{{\textbf{"No"}}}\\
\hline

\cellcolor{gray!20}\textbf{qwen2.5-7B-Ins} &
\cellcolor{gray!20}\textbf{Search-o1-Dense} &
\cellcolor{gray!20}\textbf{Search-R1} &
\cellcolor{gray!20}\textbf{Graph-R1-HippoRAG2} \\
\hline

\noalign{\renewcommand{\arraystretch}{1.35}}
\cellcolor{gray!20}\parbox[t][6cm][c]{1.5cm}{\centering\textbf{Generation}} &
\parbox[t][6cm][c]{4cm}{\centering
To answer the question of whether the Laleli Mosque and Esma Sultan Mansion are located in the same neighborhood, I need to find the locations of both structures.Assistant: \textcolor{orange}{\texttt{<|begin\_search\_query|>}} \\ location of Laleli Mosque
\textcolor{orange}{\texttt{<|end\_search\_query|>}}\\ \textcolor{red}{\texttt{<|begin\_search\_result|>}} The Laleli Mosque is located in Laleli, Fatih, Istanbul, Turkey...\textcolor{red}{\texttt{<|end\_search\_result|>}}\\
Assistant: Based on the information gathered...\{No\}

} &
\parbox[t][6cm][c]{4cm}{\centering
\textcolor{blue}{\texttt{<think>}}To answer the question ``Are the Laleli Mosque and Esma Sultan Mansion located in the same neighborhood?'',
I will first reason through what I know about these two landmarks.
I know that Laleli Mosque is a significant historical site in Istanbul, Turkey.
I need more information about the location of Esma Sultan Mansion to determine if it shares a neighborhood with Laleli Mosque.

\textcolor{blue}{\texttt{</think>}}
\textcolor{orange}{\texttt{<search>}} ``Esma Sultan Mansion location Istanbul Turkey'' \textcolor{orange}{\texttt{</search>}}

\textcolor{red}{\texttt{<information>}} >Doc 1(Title: ``Esma Sultan Mansion''):\ldots\ \textcolor{red}{\texttt{</information>}}
\textcolor{blue}{\texttt{<think>}}\ldots

Based on my search results: Laleli Mosque is in Laleli neighborhood. Esma Sultan Mansion is in Ortak\"oy neighborhood.

Since Laleli Mosque is in Laleli neighborhood and Esma Sultan Mansion is in Ortak\"oy neighborhood,
they are not located in the same neighborhood.
\textcolor{blue}{\texttt{</think>}}

\textcolor{green}{\texttt{<answer>}} No \textcolor{green}{\texttt{</answer>}}
} &
\parbox[t][6cm][c]{4cm}{\centering
\textcolor{blue}{\texttt{<think>}}
To answer this question, I need to determine the neighborhoods where the Laleli Mosque and Esma Sultan Mansion are located.
I will first search for the location of the Laleli Mosque and then for the Esma Sultan Mansion, and finally compare their neighborhoods.
\textcolor{blue}{\texttt{</think>}}

\textcolor{orange}{\texttt{<query>}}
\texttt{\{"query": "Laleli Mosque neighborhood"\}}
\textcolor{orange}{\texttt{</query>}}

\textcolor{red}{\texttt{<knowledge>}}
Wikipedia Title: Laleli Mosque\ldots
\textcolor{red}{\texttt{</knowledge>}}

\textcolor{blue}{\texttt{<think>}}
According to the Wikipedia information retrieved, the Laleli Mosque is located in Laleli, Fatih, Istanbul.
Now, I need to find the neighborhood where Esma Sultan Mansion is located.\\
\textcolor{blue}{\texttt{</think>}}
\ldots

\textcolor{green}{\texttt{<answer>}} No \textcolor{green}{\texttt{</answer>}}
} \\
\hline

\end{tabular}
} 
\end{table*}

\section{Templates}
\subsection{Instructions for Search-o1}
\subsubsection{Instruction for Search-o1}
\begin{figure*}[t]
\centering
\begin{tcolorbox}[mygreenbox, width=\textwidth, title={Instruction for Search-o1},    fonttitle=\Large\bfseries]
You are a reasoning assistant with the ability to perform web searches to help you answer the user’s question
accurately. \\
You have special tools:\\
To perform a search: write \texttt{\textless|begin\_search\_query|\textgreater} your query here \texttt{\textless|end\_search\_query|\textgreater}.\\
Then, the system will search and analyze relevant web pages, then provide you with helpful information in the
format \texttt{\textless|begin\_search\_result|\textgreater}
...search results... \texttt{\textless|end\_search\_result|\textgreater}.\\
You can repeat the search process multiple times if necessary. The maximum number of search attempts is
limited to \texttt{\{MAX\_SEARCH\_LIMIT\}}.\\
Once you have all the information you need, continue your reasoning.\\
Example:\\
Question: ``...''\\
Assistant thinking steps:\\
- I might need to look up details about ...\\
Assistant:\\
\texttt{\textless|begin\_search\_query|\textgreater}...\texttt{\textless|end\_search\_query|\textgreater}
(System returns processed information from relevant web pages)\\
Assistant continues reasoning with the new information...\\
Remember:\\
- Use \texttt{\textless|begin\_search\_query|\textgreater} to request a web search and end with \texttt{\textless|end\_search\_query|\textgreater}.\\
- When done searching, continue your reasoning.\\
Question:\{question\}
\end{tcolorbox}
\caption{Instruction for Search-o1}
\end{figure*}

\subsubsection{Instruction for Reason-in-Documents}
\begin{figure*}[t]
\centering
\begin{tcolorbox}[mygreenbox, width=\textwidth, title={Instruction for Reason-in-Documents}, fonttitle=\Large\bfseries]
Task Instruction:\\
You are tasked with reading and analyzing web pages based on the following inputs: Previous Reasoning Steps, Current Search Query, and Searched Web Pages. Your objective is to extract relevant and helpful information for Current Search Query from the Searched Web Pages and seamlessly integrate this information into the Previous Reasoning Steps to continue reasoning for the original question.\\
Guidelines:\\
1. Analyze the Searched Web Pages:\\
- Carefully review the content of each searched web page.\\
- Identify factual information that is relevant to the Current Search Query and can aid in the reasoning process for the original question.\\
2. Extract Relevant Information:\\
-Select the information from the Searched Web Pages that directly contributes to advancing the Previous Reasoning Steps.\\
-Ensure that the extracted information is accurate and relevant.\\
3. Output Format:\\
- If the web pages provide helpful information for current search query: Present the information beginning with `Final Information' as shown below.\\
Final Information\\
{[Helpful information]}\\
- If the web pages do not provide any helpful information for current search query: Output the following text.\\
Final Information\\
No helpful information found.
Inputs:\\
- Previous Reasoning Steps: \\
\texttt{\{prev\_reasoning\}}\\
- Current Search Query: \\
\texttt{\{search\_query\}}\\
- Searched Web Pages: \\
\texttt{\{document\}}\\
Now you should analyze each web page and find helpful information based on the current search query ``\texttt{\{search\_query\}}'' and previous reasoning steps.
\end{tcolorbox}
\caption{Instruction for Reason-in-Documents}
\end{figure*}

\subsubsection{Instruction for Open-Domain QA Tasks}
\begin{figure*}[h]
\centering
\begin{tcolorbox}[mygreenbox, width=\textwidth, title={Instruction for Open-Domain QA Tasks}, fonttitle=\Large\bfseries]
Please answer the following question.\\
You should provide your final answer in the format \texttt{\textbackslash boxed\{YOUR\_ANSWER\}}.\\
Question:\\
\texttt{\{question\}}
\end{tcolorbox}
\caption{Instruction for Open-Domain QA Tasks}
\end{figure*}

\subsubsection{Additional Notes}

\paragraph{\textbf{Prompting Details}}
For all the instructions above, we input them as user prompts, not system prompts. For non-reasoning models like Qwen2.5-32B-Instruct and Qwen2.5-7B-Instruct, we add a Chain-of-Thought prompt ``You should think step by step to solve it.'' before the question to explicitly make these models reason before giving the final answer.

\paragraph{\noindent\textbf{Implementation Note}}
While the instructions prompt the model to perform ``web searches'', our experiments replaced the actual Bing Web Search API with a retrieval server. The model's generated search queries were intercepted and redirected to our retrieval corpus. The retrieved knowledge chunks were then formatted to mimic web search results before being returned to the model.\\

\subsection{GraphSearch Instruction}
\subsubsection{Query Decomposition}
\begin{figure*}[t]
\centering
\begin{tcolorbox}[mybluebox, width=\textwidth, title={Query Decomposition}]
---Role---\\
You are a helpful assistant specializing in complex query decomposition.\\[2pt]
---Goal---\\
Given a main query, your task is to break it down into several atomic sub-queries, which should
directly correspond to parts of the original query.\\[2pt]
---Instructions---\\
- Decompose the main query into clear and actionable sub-queries that represent smaller, solvable
pieces of the main question.\\
- Ensure that each sub-query addresses one specific entity or concept, with the goal of retrieving
information that will answer the overall main query.\\
- Use sequential numbering (i.e., \texttt{\#1}, \texttt{\#2}, etc.) to represent answers of previous sub-queries. For
example, \texttt{\#1} refers to the answer of Sub-query 1.\\
- Make sure the sub-queries are logically ordered, where the output of one sub-query might feed into
the next.\\
- The final output should be in JSON format, where each sub-query is listed as a key-value pair.\\[2pt]
---Examples---\\
Main Query:\\
How many times did plague occur in the place where the creator of The Worship of Venus died?\\
Sub-queries:\\
\{\{\\
"Sub-query 1": "Who is the creator of The Worship of Venus?",\\
"Sub-query 2": "Where did \texttt{\#1} die?",\\
"Sub-query 3": "How many times did plague occur in \texttt{\#2}?"\\
\}\}\\[4pt]
Main Query:\\
When did the city where Hillcrest High School is located become the capital of the state where the
screenwriter of The Poor Boob was born?\\
Sub-queries:\\
\{\{\\
"Sub-query 1": "Where is Hillcrest High School located?",\\
"Sub-query 2": "Who is the screenwriter of The Poor Boob?",\\
"Sub-query 3": "Where was \texttt{\#2} born?",\\
"Sub-query 4": "When did the city from \texttt{\#1} become the capital of the state from \texttt{\#3}?"\\
\}\}\\[4pt]
Main Query:\\
What crop, which is a big feeder of nitrogen, has a gross income of \$1,363.00 per acre and a net profit
of \$658.00?\\
Sub-queries:\\
\{\{\\
"Sub-query 1": "Which crops are considered big feeders of nitrogen?",\\
"Sub-query 2": "Among \texttt{\#1}, which crop has a gross income of \$1,363.00 per acre?",\\
"Sub-query 3": "Does \texttt{\#2} have a net profit of \$658.00?"\\
\}\}\\[4pt]
---Input---\\
Main Query:\\
\{query\}\\[2pt]
---Output---

\end{tcolorbox}
\caption{GraphSearch Query Decomposition}
\end{figure*}

\subsubsection{Query Decomposition (Knowledge Graph)}
\begin{figure*}[t]
\centering
\begin{tcolorbox}[mybluebox, width=\textwidth, title={Query Decomposition (Knowledge Graph)}]
---Role---\\
You are a helpful assistant specializing in complex query decomposition for knowledge graph retrieval.\\[2pt]
---Goal---\\
Given a main query, your task is to break it down into atomic sub-queries in the form of subject-predicate-object triples. These should correspond directly to parts of the original query and be suitable for querying a knowledge graph.\\[2pt]
---Instructions---\\
- Decompose the main query into a sequence of sub-queries, where each sub-query consists of one or more atomic triples in the format: ("entity1", "relationship", "entity2").\\
- Replace any unknown entity with a placeholder such as Entity\texttt{\#1}, Entity\texttt{\#2}, etc.\\
- Maintain logical ordering, where the result of one sub-query (e.g., Entity\texttt{\#1}) might be required for the next.\\
- Each sub-query may contain more than one triple if needed to express the full meaning.\\
- The final output should be in JSON format, where each key is a sub-query and the value is a list of atomic triples enclosed in parentheses.\\[2pt]
---Examples---\\
Main Query:\\
How many times did plague occur in the place where the creator of The Worship of Venus died?\\
Sub-queries:\\
\{\{\\
"Sub-query 1": [("The Worship of Venus", "is created by", "Entity\texttt{\#1}")],\\
"Sub-query 2": [("Entity\texttt{\#1}", "died at", "Entity\texttt{\#2}")],\\
"Sub-query 3": [\\
("Plague", "occur in", "Entity\texttt{\#2}"),\\
("Plague", "times of occur", "Entity\texttt{\#3}")\\
]\\
\}\}\\[4pt]
Main Query:\\
When did the city where Hillcrest High School is located become the capital of the state where the screenwriter of The Poor Boob was born?\\
Sub-queries:\\
\{\{\\
"Sub-query 1": [("Hillcrest High School", "is located in", "Entity\texttt{\#1}")],\\
"Sub-query 2": [("The Poor Boob", "has screenwriter", "Entity\texttt{\#2}")],\\
"Sub-query 3": [("Entity\texttt{\#2}", "was born in", "Entity\texttt{\#3}")],\\
"Sub-query 4": [\\
("Entity\texttt{\#1}", "is capital of", "Entity\texttt{\#3}"),\\
("Entity\texttt{\#1}", "became capital at", "Entity\texttt{\#4}")\\
]\\
\}\}\\[4pt]
Main Query:\\
What crop, which is a big feeder of nitrogen, has a gross income of \$1,363.00 per acre and a net profit of \$658.00?\\
Sub-queries:\\
\{\{\\
"Sub-query 1": [("Entity\texttt{\#1}", "is a", "crop that is a heavy nitrogen feeder")],\\
"Sub-query 2": [("Entity\texttt{\#1}", "has gross income per acre", "\$1,363.00")],\\
"Sub-query 3": [("Entity\texttt{\#1}", "has net profit", "\$658.00")]\\
\}\}\\[4pt]
---Input---\\
Main Query:\\
\{query\}\\[2pt]
---Output---
\end{tcolorbox}
\caption{GraphSearch Query Decomposition (KG)}
\end{figure*}

\subsubsection{Evidence Verification}
\begin{figure*}[t]
\centering
\begin{tcolorbox}[mybluebox, width=\textwidth, title={Evidence Verification}]
---Role---\\
You are a critical evaluator specializing in verifying the logical soundness and evidential sufficiency of model-generated responses.\\[2pt]
---Goal---\\
Given a user query, retrieved context data, and the model-generated response, your task is to evaluate whether the response forms a rigorous
logical loop supported by the provided evidence.\\[2pt]
---Instructions---\\
- Carefully examine whether the response is \textbf{strictly grounded} in the retrieved context data.\\
- Assess whether the reasoning process forms a \textbf{complete logical chain}, without missing steps or unsupported leaps.\\
- Identify if there are \textbf{evidence gaps, low-confidence claims, or speculative statements}.\\
- If the response demonstrates a well-supported, confident, and logically closed argument, conclude your analysis with \textbf{``Yes''}.\\
- If the response shows hesitation, incomplete reasoning, or lacks solid evidence support, conclude your analysis with \textbf{``No''}.\\[2pt]
---Input---\\
User-Query:\\
\{query\}\\
Retrieved Context Data:\\
\{context\_data\}\\
Model Response:\\
\{model\_response\}\\[2pt]
---Output---
\end{tcolorbox}
\caption{GraphSearch Evidence Verification}
\end{figure*}

\subsubsection{Deep Answer Generation}
\begin{figure*}[t]
\centering
\begin{tcolorbox}[mybluebox, width=\textwidth, title={Deep Answer Generation}]
---Role---\\
You are a helpful assistant specializing in complex question answering.\\[2pt]
---Goal---\\
Given a complex query and retrieved context data, your task is to construct a logically sound, step-by-step answer. 
Your explanation should follow a rigorous reasoning path, incorporate relevant evidence, and establish clear relationships between the entities.\\[2pt]
---Instructions---\\
- Break down the reasoning process into clear, coherent steps.\\
- Use context data explicitly to support each reasoning step.\\
- Make sure relationships between entities are logically explained.\\[2pt]
---Input---\\
Query:\\
\{query\}\\
Context Data:\\
\{context\_data\}\\[2pt]
---Output---
\end{tcolorbox}
\caption{GraphSearch Deep Answer Generation}
\end{figure*}

\subsubsection{Query Expansion}
\begin{figure*}[t]
\centering
\begin{tcolorbox}[mybluebox, width=\textwidth, title={Query Expansion}]
---Role---\\
You are a helpful assistant specializing in query expansion for evidence completion.\\[2pt]
---Goal---\\
Given a main query, retrieved context data, the model-generated response, and the evidence verification analysis, your task is to perform \textbf{query expansion}.\\
If the evidence verification analysis shows that the current evidence is insufficient to support the logical chain of the response, generate one or more additional sub-queries.\\
These sub-queries should aim to cover missing retrieval scenarios, fill in the evidence gaps, and guide towards a more complete and confident logical reasoning chain.\\[2pt]
---Instructions---\\
- Use the retrieved context data, especially any existing sub-queries in the retrieval history, as references when generating new sub-queries.\\
- Focus on producing \textbf{complementary sub-queries} that address aspects not yet fully supported by evidence.\\
- Avoid duplicating existing sub-queries; instead, expand into related but uncovered areas.\\
- Keep sub-queries clear, specific, and directly actionable for retrieval.\\
- Output should be in the form of a \textbf{Python-style List of strings}, where each string is a new sub-query.\\[2pt]
---Input---\\
Main Query:\\
\{query\}\\
Retrieved Context Data:\\
\{context\_data\}\\
Model Response:\\
\{model\_response\}\\
Evidence Verification Analysis:\\
\{evidence\_verification\}\\[2pt]
---Output---
\end{tcolorbox}
\caption{GraphSearch Query Expansion}
\end{figure*}

\subsection{Search-R1 Instruction}
\begin{figure*}[t]
\centering
\begin{tcolorbox}[mygraybox, width=\textwidth, title={Search-R1 Instruction}]
Answer the given question.

You must conduct reasoning inside <think> and </think> first every time you get new information.

After reasoning, if you find you lack some knowledge, you can call a search engine by <search> query </search> and it will return the top searched results between <information> and </information>.

You can search as many times as your want.

If you find no further external knowledge needed, you can directly provide the answer inside <answer> and </answer>, without detailed illustrations. For example, <answer> Beijing </answer>. 

Question: \{question\}
\end{tcolorbox}
\caption{Search-R1 Instruction}
\end{figure*}

\subsection{Graph-R1 Instruction}
\begin{figure*}[t]
\centering
\begin{tcolorbox}[mypurplebox, width=\textwidth, title={Graph-R1 Instruction},    fonttitle=\Large\bfseries]
Answer the given question. \\
You can query from knowledge base provided to you to answer the question. \\
You can query knowledge as many times as you want.
You must first conduct reasoning inside <think>...</think>. \\
If you need to query knowledge, you can set a query statement between <query>...</query> to query from knowledge base after <think>...</think>. \\
When you have the final answer, you can output the answer inside <answer>...</answer>. \\
Output format for query: \\
<think>\\
... \\
</think>\\
<query>\\
... \\
</query> \\
Output format for answer:\\
<think>\\
... \\
</think>\\
<answer>\\
... \\
</answer> \\
Question: \{question\}
\end{tcolorbox}
\caption{Graph-R1 Instruction}
\end{figure*}

\end{document}